\documentclass[twocolumn]{aastex631}                 %for arXiv
\usepackage{amsmath}
\usepackage{url}
\shortauthors{Mintz et al.}
\begin{document}

% REU: The title is the single most important element of the paper, because it is the part
% most widely used for literature searches.  
%
% Your title should be about  WHAT you did, and not tightly focused on HOW you did it.  
% The methods you used are of secondary importance to what you accomplished.  
% In other words, put your science first.  Try not to use the overused word "study" in the title.

\title{Testing Models of Triggered Star Formation with Young Stellar Objects in Cepheus~OB4}

\correspondingauthor{Abby\ Mintz}
\email{abby.mintz@yale.edu}

% REU: This shows how to format the names.  Use your full name, spelled out the way
% you want it to appear in the literature.  The 16-digit number inside the [] should
% be replaced with your ORCID.  If you don't have one, you should sign up for one,
% and put that number here and in all your future papers.  
% For your affiliation, use your home institution and address, not the CfA.
% You and your advisor should work out together whose names appear here,
% the basic criterion is that people who have made essential contributions 
% to the manuscript should be included.

\author[0000-0002-9816-9300]{Abby Mintz}
\affil{Department of Astronomy, Yale University,
52 Hillhouse Ave.,
New Haven, CT 06511, USA}
\affiliation{Center for Astrophysics $|$ Harvard \& Smithsonian,
60 Garden St., MS-65,
Cambridge MA 02138, USA}

\author[0000-0002-5599-4650]{Joseph L. Hora}
\affiliation{Center for Astrophysics $|$ Harvard \& Smithsonian,
60 Garden St., MS-65,
Cambridge MA 02138, USA}

\author[0000-0001-9065-6633]{Elaine Winston}
\affiliation{Center for Astrophysics $|$ Harvard \& Smithsonian,
60 Garden St., MS-65,
Cambridge MA 02138, USA}

% \collaboration{(AAS Journals Data Scientists collaboration)}

% Mark off the abstract in the ``abstract'' environment. 
\begin{abstract}

OB associations are home to newly formed massive stars, whose turbulent winds and ionizing flux create \ion{H}{2} regions rich with star formation. Studying the distribution and abundance of young stellar objects (YSOs) in these ionized bubbles can provide essential insight into the physical processes that shape their formation, allowing us to test competing models of star formation. In this work, we examined one such OB association, Cepheus~OB4 (Cep~OB4) – a well-suited region for studying YSOs due to its Galactic location, proximity, and geometry. We created a photometric catalog from \textit{Spitzer}/IRAC mosaics in  bands 1 (3.6~\micron) and 2 (4.5~\micron). We supplemented the catalog with photometry from \textit{WISE}, 2MASS, IRAC bands 3 (5.8~\micron) and 4 (8.0 \micron), MIPS 24~\micron, and MMIRS near IR data. We used color-color selections to identify 821 YSOs, which we classified using the IR slope of the YSOs' spectral energy distributions (SEDs), finding 67 Class I, 103 flat spectrum, 569 Class II, and 82 Class III YSOs. We conducted a clustering analysis of the Cep~OB4 YSOs and fit their SEDs.  We found many young Class I objects distributed in the surrounding shell and pillars as well as a relative age gradient of unclustered sources, with YSOs generally decreasing in age with distance from the central cluster. Both of these results indicate that the expansion of the \ion{H}{2} region may have triggered star formation in Cep~OB4.

\end{abstract}



\section{Introduction} \label{sec:intro}

Many stars in our galaxy formed in OB associations, loose unbound systems of massive O and B stars \citep{Briceno2007, Lada1999}. The environments of OB associations, typically embedded in giant molecular clouds (GMCs), are strongly affected by massive stars ($>8$ M$_\odot$) whose turbulent and fast-paced lives dramatically shape their surroundings \citep{Zinnecker2007}. Such OB associations also host less massive star formation and therefore young, less massive stars. 
Massive stars are an integral part of galactic evolution, providing heavy metals, massive outflows, and turbulent winds that greatly influence their environments \citep{Zinnecker2007}. Despite their important role in shaping astrophysical conditions, the processes that lead to their formation are still not entirely understood. Directly observing O and B stars in the midst of formation is difficult due to dust extinction, their relatively low numbers and greater distances, and the relative brevity of their formation period. 

\begin{figure*}
\begin{center}
\includegraphics[width=\linewidth,angle=0]{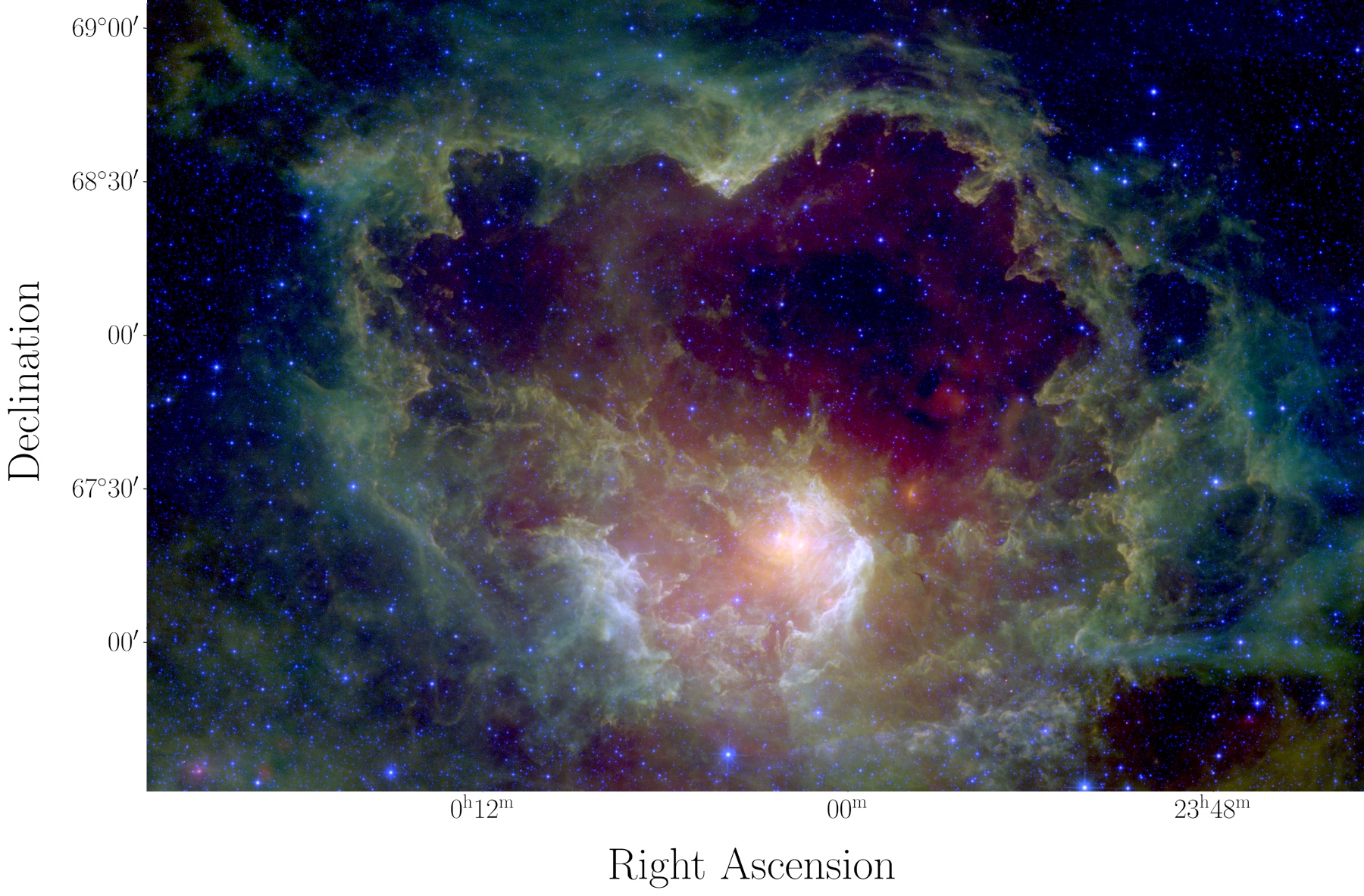}
\caption{A color image of the Cep~OB4 field using the \textit{WISE} 3.4, 12, and 22~\micron\ images (blue: 3.4~\micron, green: 12~\micron, red: 22~\micron).  Berkeley~59 – the cluster of early-type stars – can be seen in the southern central region of the image. The shell of the molecular cloud surrounds most of the bright stars in the image. Pillars  around the edges of the shell point in towards the OB association.\label{fig:image}}
\end{center}
\end{figure*}

Observing their effects on less-massive stars' presence and development in OB associations is more feasible, in part due to the creation of ionized \ion{H}{2} regions and the longer life cycle of less-massive stars. The stellar winds and ionizing flux from massive stars form \ion{H}{2} regions within GMCs, pushing material from inside the OB association to the outskirts of the \ion{H}{2} regions. These ionized bubbles have borders of dense material, nebulous features, and filaments, all of which can be sites of lower-mass star formation. Many star-forming \ion{H}{2} regions with bubble morphologies have been studied previously, including Sh2-236 \citep{Ortega2020}, Sh2-48 \citep{Ortega2013}, and Vela~OB2 \citep{Cantat2019} among others \citep{Xu2014}.

The presence, distribution, and location of YSOs provides information about recent star formation in the OB associations and can help determine what physical processes are at play. The collect and collapse model posits that winds and supernovae caused by massive stars can compress material sufficiently to trigger star formation by spurring gravitational collapse \citep{Elmegreen1977}. This process would lead to the presence of YSOs around the outskirts of the \ion{H}{2} regions, decreasing in age with increasing distance from the OB association's center. A second model, radiation-driven implosion, suggests that star formation may result from the ionization front's interaction with dense molecular regions, triggering increased density and collapse \citep{Reipurth1983}. This model would yield clusters of YSOs in molecular structures like pillars that the \ion{H}{2} region has expanded to include. A third proposed mechanism involves molecular cloud collisions \citep{Loren1976}. YSOs would be formed in the dense regions created by the colliding material and would therefore be found in the areas of collision. 

The OB association Cepheus~OB4 (Cep~OB4) is a potentially useful testing ground for star formation models due to its Galactic location, proximity, and geometry. Cep~OB4 was first identified in 1959 along with an interior cluster of early-type stars – Berkeley~59 \citep[Be~59;][]{Blanco1959, Kun2008}, which has been recently studied by \citet{Rosvick2013} and is estimated to be $\sim$ 2~Myr old \citep{Panwar2018}. Several OB stars in the region have been cataloged, including some in the central Be~59 area \citep{Skiff2014}. These massive stars are likely the source of the shockwave that evacuated the molecular material and created the bubble's shape and structure. Cep~OB4's Galactic latitude of about $+5$\degr\ places it off of the densest area of the Galactic plane, which results in lower foreground and background contamination. At $\sim$1.1~kpc \citep{Kuhn2019} it is relatively close for an OB association, allowing for better resolution of individual sources and detection of lower-mass YSOs – down to a few tenths of a solar mass \citep{Hora2018}. With an estimated radius of 140~pc \citep[][adjusted to the 1.1~kpc distance]{Olano2006}, it is larger than the typical OB association where the average diameter has been estimated to be $\sim$140~pc \citep{Garmany1994}. Be~59 is also embedded in and expanding into an approximately circular molecular cloud with a diameter of $\sim$80~pc, allowing us to test models of sequential star formation.

In this work, we use new mosaics of the Cep~OB4 OB association taken with the Infrared Array Camera \citep[IRAC;][]{Fazio2004} on the \textit{Spitzer Space Telescope} \citep{Werner2004} to identify YSOs based on their excess IR emission (a result of the reprocessed stellar radiation in their surrounding dusty material). In Section~\ref{sec:obs} we describe the images and photometric catalogs used in our analysis. In Section~\ref{sec:red}, we perform photometry on the new images and cross match the resulting catalog with additional near and mid infrared photometry to extend our wavelength coverage.  

In Section~\ref{sec:YSO} we use the compiled photometric catalog to remove background sources, identify YSOs with various color-color selections, and classify the YSOs based on the slopes of their spectral energy distributions (SEDs). In Section~\ref{sec:distribution} we study the spatial clustering and distribution of the YSOs, focusing especially on the youngest objects. In Section~\ref{sec:sed}, we fit the YSO SEDs to determine their relative masses, which we used to construct the initial mass function for Cep~OB4. In Section~\ref{sec:results} we discuss our results, comparing our YSO analysis to various models of triggered star formation. We conclude with a summary of our findings in Section~\ref{sec:conclusion}.

\begin{deluxetable}{rcrl}
\tablecaption{IRAC observations used in mosaics\label{tab:aors}}
\tablewidth{0pt}
\tablehead{
\colhead{AOR} & \colhead{Observation Date} & \colhead{Program} & \colhead{PI\tablenotemark{a}}
}
\startdata
3658240 & 2003-12-23 13:33:58 & 6 & Fazio\\
6034688 & 2004-07-28 16:01:13 & 202 & Fazio\\
6034176 & 2004-07-28 16:06:27 & 202 & Fazio\\
41713920 & 2011-01-28 08:46:17 & 70062 & Kirkpatrick\\
44743424 & 2012-04-19 10:52:50 & 80109 & Kirkpatrick\\
48001280 & 2013-04-12 10:27:38 & 90179 & Getman\\
68418048 & 2018-11-17 11:04:31 & 14005 & Hora\\
68418816 & 2018-11-17 12:12:22 & 14005 & Hora\\
68418560 & 2018-11-17 14:25:05 & 14005 & Hora\\
68419072 & 2018-11-18 04:07:57 & 14005 & Hora\\
68418304 & 2018-11-19 09:37:45 & 14005 & Hora\\
68516608 & 2019-01-04 06:17:35 & 14005 & Hora\\
68515840 & 2019-01-04 07:24:52 & 14005 & Hora\\
68516096 & 2019-01-05 18:59:24 & 14005 & Hora\\
68516352 & 2019-01-05 21:01:06 & 14005 & Hora\\
68515584 & 2019-01-05 22:16:52 & 14005 & Hora\\
68515328 & 2019-01-06 08:18:10 & 14005 & Hora\\
68516864 & 2019-01-06 10:25:56 & 14005 & Hora\\
\enddata
\tablenotetext{a}{The programs listed here are \citet[][Program 6]{Fazio2004b}, \citet[][Program 202]{Fazio2004c}, \citet[][Program 70062]{Kirkpatrick2010}, \citet[][Program 80109]{Kirkpatrick2011}, \citet{Getman2012}, and \citet[][Program 14005]{Hora2018}.}
\end{deluxetable}

\vspace{-1em}
\section{Observations} \label{sec:obs}
\subsection{Spitzer IRAC} \label{subsec:spitzer}
Cep~OB4 was mapped by several \textit{Spitzer} programs during the mission. The list of Astronomical Observation Requests (AORs) used to construct our mosaics is shown in Table~\ref{tab:aors}. The programs prior to 2013 obtained individual pointings or small maps covering sources of special interest in the region. The programs conducted in 2003 and 2004 obtained images in all four IRAC bands (3.6, 4.5, 5.8, and 8~\micron). The other programs were conducted during the \textit{Spitzer} Warm Mission and therefore obtained data only in the 3.6 and 4.5~\micron\ IRAC bands. Program 90179 mapped an approximately 1\degr$\times$1\degr\ area covering the Be~59 cluster using 30 second HDR mode (1.2 and 30 second frames) with 5 dithers per map pointing. The majority of the mosaic was obtained in Program 14005 using 12 second HDR mode (0.6 and 12 second frames) and 3 dithers per pointing. Several AORs were necessary to map the full area.

We constructed the  mosaic images using the corrected Basic Calibrated Data (cBCD) products produced by the \textit{Spitzer} Science Center pipeline. In addition to the basic pipeline calibration and artifact correction, we constructed and subtracted  residual background frames for each AOR to correct for any remaining background structure. We also used the artifact correction software \texttt{imclean} v3.2 \citep{Hora2021} to correct the remaining column pulldown and banding effects that occur for bright stars. We used the IRACproc mosaicking software \citep{Schuster2006} to combine the data into one ``long frame" (using the 12 and 30 second images) and one ``short frame" (using the 0.6 and 1.2 second images) mosaic for each band. A color image of the final long frame 3.6~\micron \ mosaic (and the \textit{WISE} 12 and 22~\micron \ images) is shown in Figure~\ref{fig:image}.
 
\subsection{MIPS}
Several areas of special interest were imaged previously by \textit{Spitzer}'s Multi-Band Imaging Photometer  \citep[MIPS;][]{Rieke2004} at 24~\micron\ in Programs 6 and 202 (see Table~\ref{tab:aors}). We obtained the post-BCD level 2 mosaics from the \textit{Spitzer} Science Center. 
\
\subsection{WISE} \label{subsec:WISE}

The \textit{Wide-field Infrared Survey Explorer} \citep[\textit{WISE};][]{Wright2010} mission surveyed the entire sky in four infrared bands: 3.4, 4.6, 12, and 22~\micron. While the \textit{WISE} images allowed us to detect IR emission and identify molecular clouds and massive stars, \textit{WISE}'s resolution ($\sim$6\arcsec\ at 3.4~\micron) limits its ability to resolve the low-mass young stars in this region, which are typically found in spatially dense clusters and can result in source confusion. We therefore relied primarily on our IRAC catalog (with a point spread function FWHM of 1\farcs8), cross matching our detected sources with \textit{WISE} to include photometry in the 12 and 22~\micron\ bands. In addition to providing higher confidence in the reliability of the cross matched sources, the supplemental photometry helped us construct the SEDs that we used to classify YSOs (see Section~\ref{sec:sed}). 

We obtained the \textit{WISE} photometry from the AllWISE Source Catalog in the NASA/IPAC Infrared Science Archive. We queried a region with $355\degr < \alpha< 7\degr$ and $66\fdg3 <\delta < 70\degr$. Because of the irregular edges of the IRAC mosaic, we chose a region that is considerably larger than the area covered by the IRAC images. Therefore, while this query resulted in a data set of 339,806 sources, only 144,280 sources were within 10\arcsec\ of an IRAC source (in the vicinity of the area covered by our IRAC mosaics).

\subsection{2MASS} \label{subsec:2MASS}

The Two Micron All Sky Survey \citep[2MASS;][]{Skrutskie2006} survey imaged nearly the entire sky in the near infrared \textit{J-} (1.25~\micron), \textit{H-} (1.65~\micron), and \textit{K$_s$-} (2.16~\micron) bands. As with \textit{WISE}, we cross matched our IRAC sources with the publicly available 2MASS photometric catalog from the NASA/IPAC Infrared Science Archive. We queried the same region described in Section~\ref{subsec:WISE} from the 2MASS All-Sky Point Source Catalog. We obtained a 2MASS data set of 310,956 sources, with 144,065 within the region covered by the IRAC mosaics.

\subsection{MMIRS Near-IR Imaging} \label{subsec:MMIRS}
\begin{figure}
\begin{center}
\includegraphics[width=0.95\linewidth,angle=0]{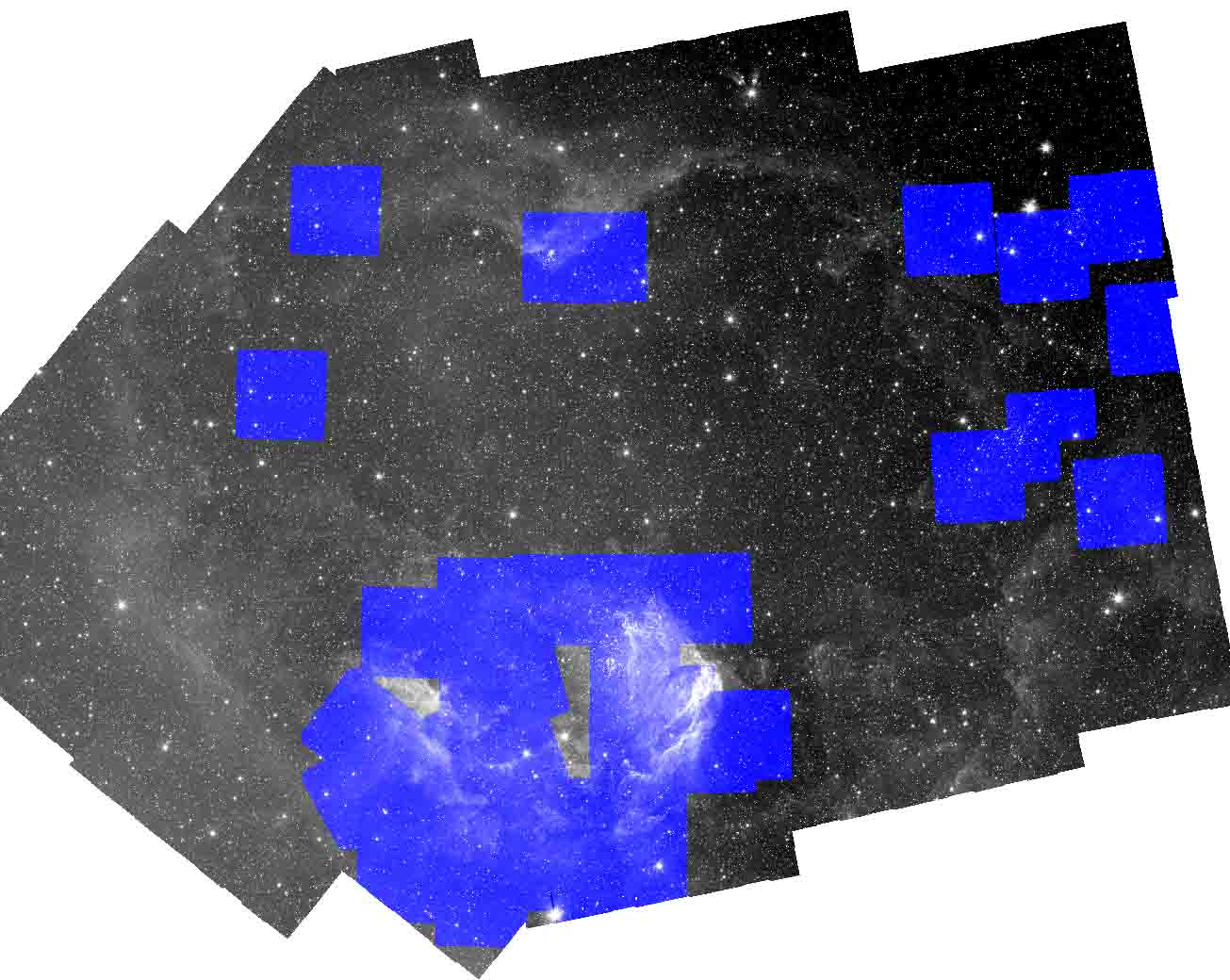}
\includegraphics[width=0.95\linewidth,angle=0]{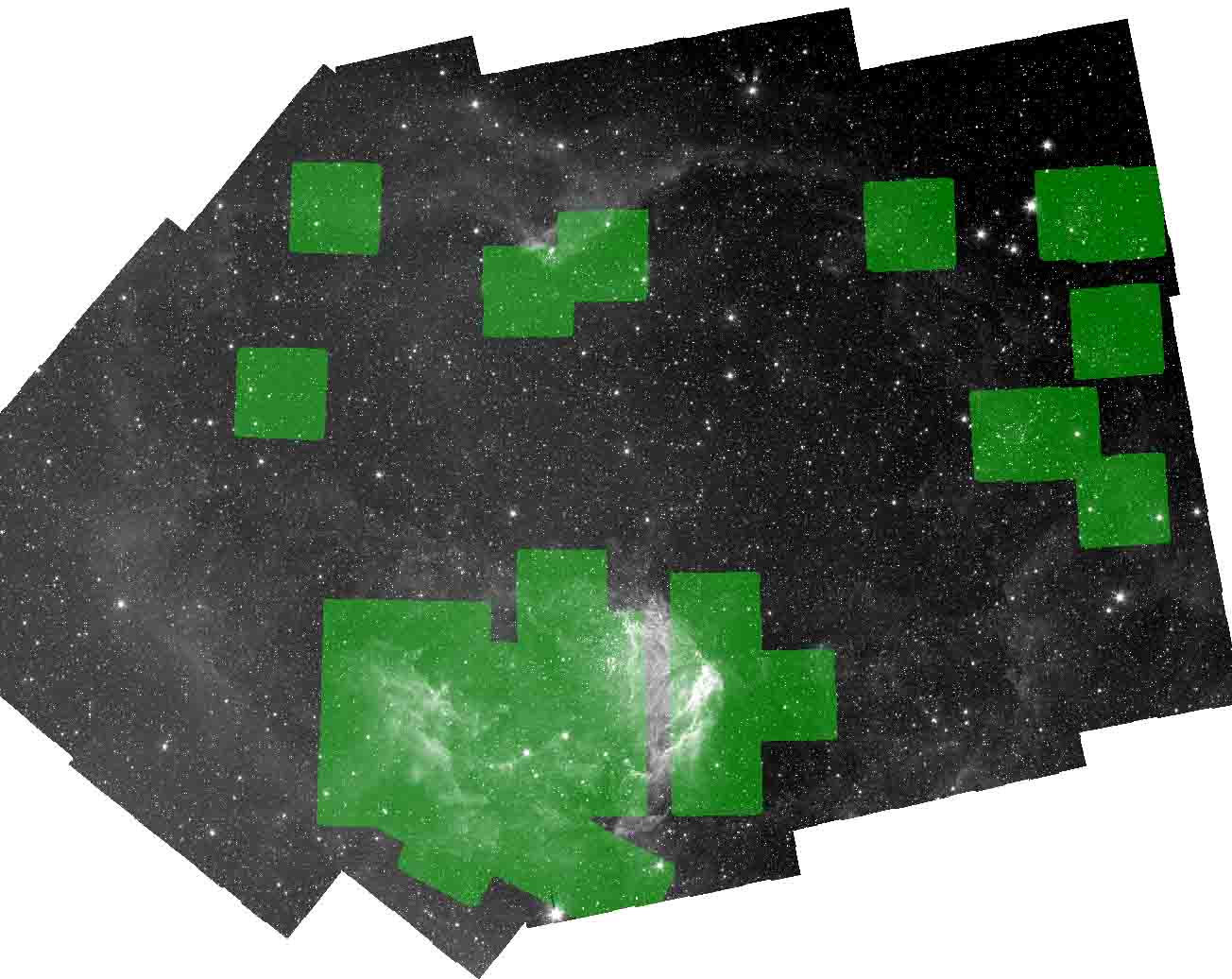}
\includegraphics[width=0.95\linewidth,angle=0]{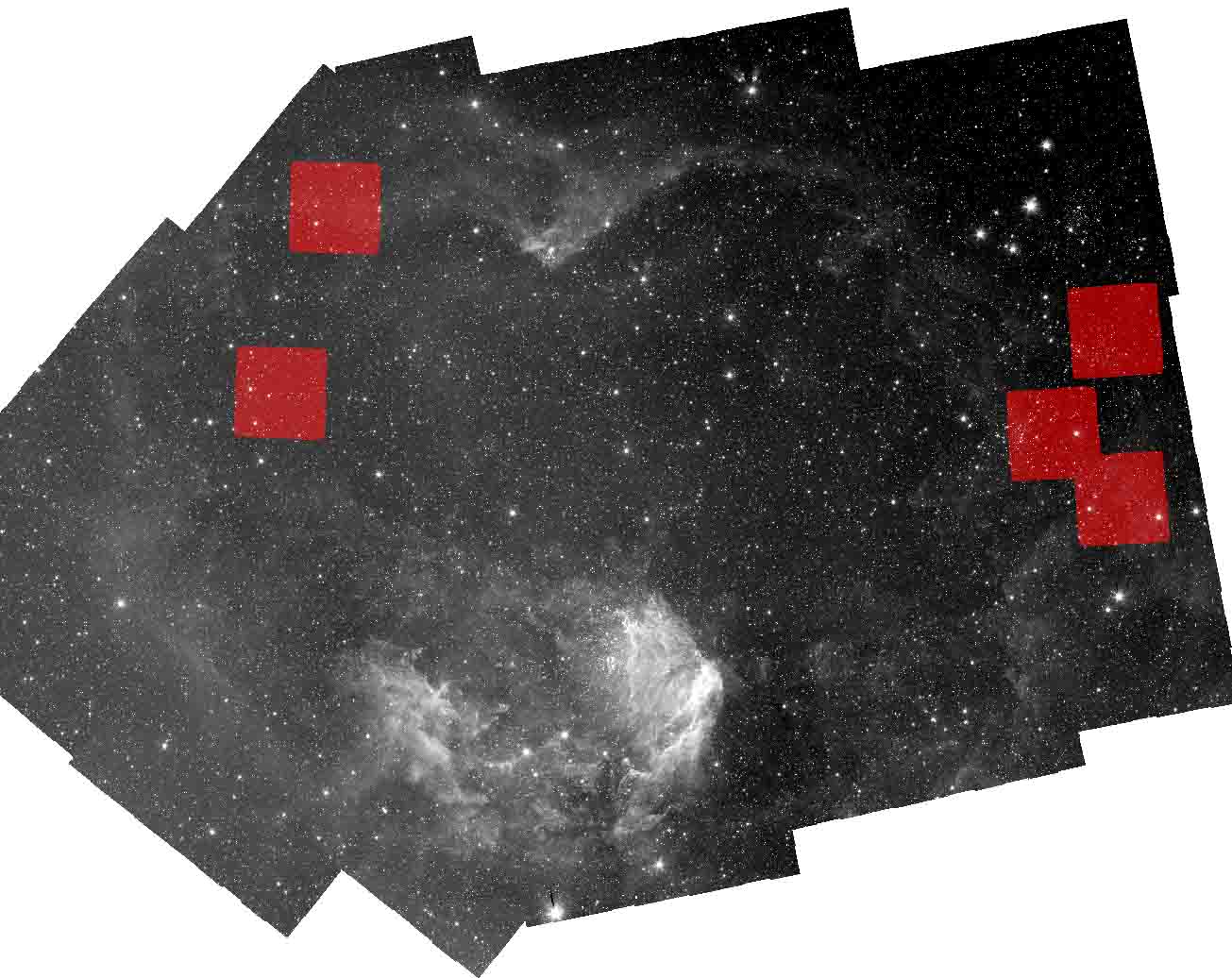}
\caption{The near-IR MMIRS coverage, plotted on the IRAC 4.5~\micron\ image of Cep~OB4. Top: the MMIRS {\it J}-band sources plotted in blue. Middle: the MMIRS {\it H}-band sources plotted in green. Bottom: the MMIRS {\it K}-band sources plotted in red. 
\label{fig:MMIRS}}
\end{center}
\end{figure}

Near-IR images were obtained at {\it J, H,} and {\it K} with the MMT and Magellan Infrared Spectrograph \citep[MMIRS;][]{McLeod2012} at the MMT on Mt. Hopkins, AZ on several nights in 2019 December - 2020 January (program SAO-8-19c) and 2020 December - 2021 January (program SAO-12-21a). The instrument field of view (FOV) is $\sim$ 6.9\arcmin$\times$6.9\arcmin\ in size, with 0\farcs2 pixels. The typical seeing-limited point source FWHM sizes were in the range 0\farcs5 -- 0\farcs7.  A large area was mapped by performing several dithered frames at each position and then offsetting by the FOV size to a new position for another set of dithered frames. The data were first processed using the MMIRS pipeline \citep{Chilin2015} to perform the sample-up-the-ramp and linearization corrections. The images were then background-subtracted and mosaicked by a custom reduction program which uses tools in the {\tt astropy} package \citep[][]{Astropy2013, Astropy2018} to align and average the frames with outlier-rejection. The combined spatial coverage in each band is shown in Figure~\ref{fig:MMIRS}.

\subsection{PanSTARRS Photometry}\label{subsec:panstarrs}
We used optical photometry  from the PanSTARRS data release 2 \citep{Flewelling2020} available on the Mikulski Archive for Space Telescopes\footnote{\url{https://archive.stsci.edu/panstarrs/}}.  The PanSTARRS archive was searched for objects within 1\arcsec\ of the positions of the YSOs identified in Section~\ref{sec:YSO}. A total of 33 objects were found to match the positions of the YSO candidates identified from the IRAC data. The {\it r} and {\it i} band optical photometry of these sources was used in the SED fitting described in Section~\ref{sec:sed}.

\section{Data Reduction}\label{sec:red}
\subsection{IRAC Photometry} \label{subsec:phot}

We performed photometry on the IRAC mosaics using {\tt Source Extractor} \citep[SExtractor;][]{Bertin1996}. The details of the source extraction and photometry are provided in Appendix \ref{appendix:ExtractPhotoIRAC}. 
\begin{figure}
\begin{center}
\includegraphics[width=\linewidth,angle=0]{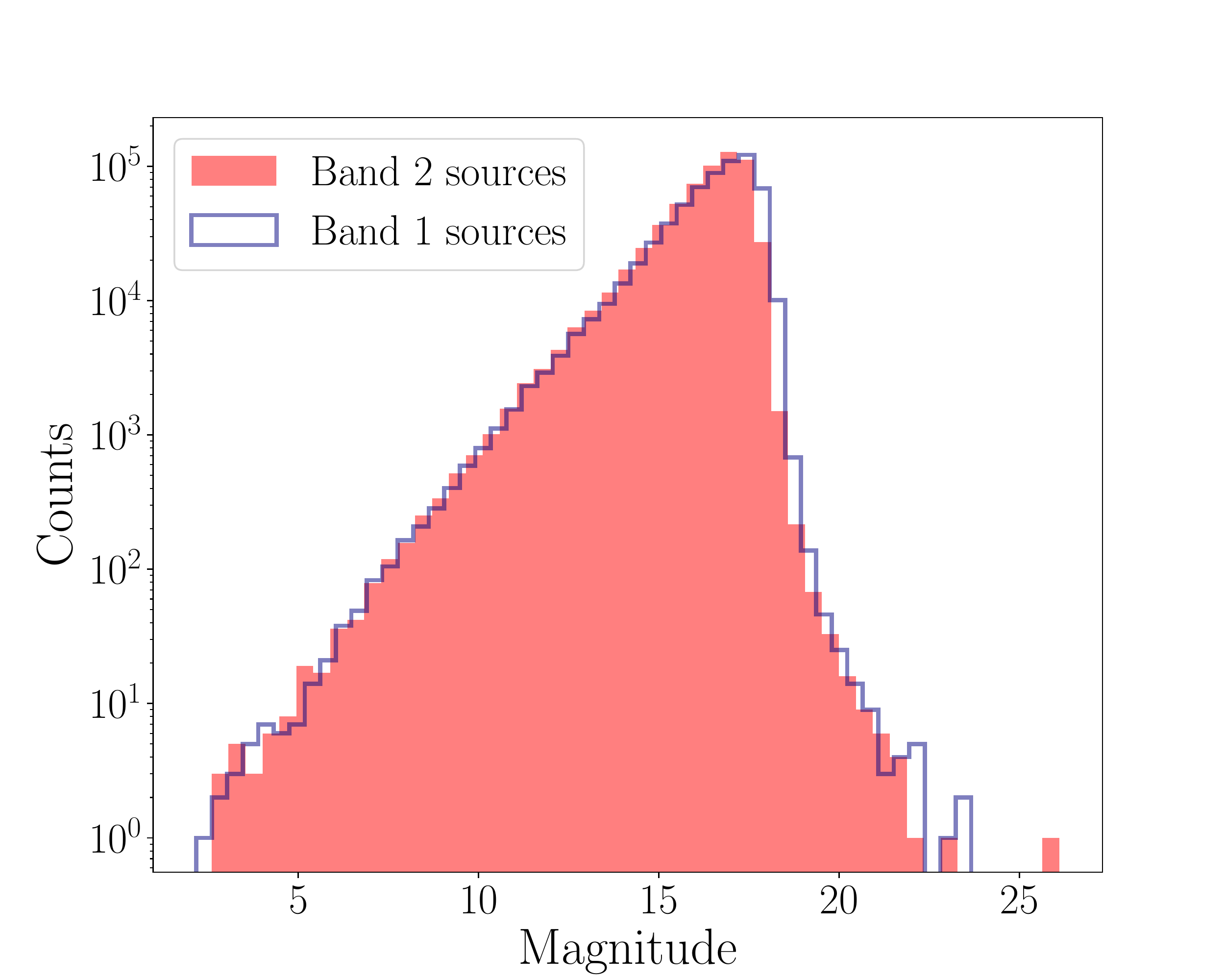}
\caption{Histograms of detected source magnitudes for IRAC bands 1 and 2, with counts plotted on a log scale. Band 1 sources are plotted in navy and band 2 sources are plotted in red. The number of sources peaks at about 17 magnitudes and drops sharply dimmer of 17 magnitudes. This drop reflects the lower limit of our detection capability. 
\label{fig:sourcehist}}
\end{center}
\end{figure}
We then matched sources in the short and long frame catalogs for each IRAC band. We used the difference in long and short frame magnitude to determine an appropriate magnitude cutoff to account for saturation in the long frame exposures. We obtained magnitude cutoffs of 10.8, 10.3, 9.5, and 9.2 for bands 1, 2, 3, and 4 respectively by identifying the brightest magnitude above which there was significant disagreement between the short and long frame magnitudes. For sources brighter than the cutoff, we replaced the magnitudes from the long frame mosaics with the values obtained from the short frame mosaic. In the case that one source had a magnitude below the cutoff and the other above (for example if the long frame magnitude was brighter than the cutoff and the short frame magnitude fainter than the cutoff), we used the short frame value. There were several bright sources in each band that were detected in only one of the two exposures (18 in band 1 and 73 in band 2, 7 in band 3, and 13 in band 4). These sources were examined manually and found to be either false detections or true detections whose positions were affected by saturation. These sources were included or excluded from the catalogs appropriately.

We performed the source extraction again using a separate set of extraction parameters for the long frame images in the Be~59 region. This region is significantly more nebulous than the rest of the field and so extracting an optimal number of sources required different background parameters. We only included sources from this Be~59 extraction that were not present in the previous extractions, adding an additional 18,818 source in band 1 and 16,296 in band 2.

After replacing saturated long frame values with the short frame photometry and adding the additional Be~59 sources, we obtained a band 1 catalog of 654,850 sources and a band 2 catalog of 616,570 sources. Histograms of the band 1 and band 2 magnitudes of detected sources are shown in Figure~\ref{fig:sourcehist}. We then matched the band 1 and band 2 sources obtained from our data using a cross matching distance of 1\farcs5 to create a single catalog of sources with magnitudes in at least one band. Our final catalog derived from our IRAC band 1 and 2 data contained 752,615 total unique sources; 69\% had measured magnitudes in both bands, 18\% had magnitudes only in band 1, and 13\% had magnitudes only in band 2.

After compiling the IRAC catalog in bands 1 and 2, we cross matched with the catalogs from bands 3 and 4. We only included band 3 and 4 sources that had corresponding sources in bands 1 or 2. We found 714 cross matched sources with photometry in bands 3 and 4, 1,437 with photometry only in band 3, 927 sources with photometry in only band 4, and 749,537 sources with photometry in neither band 3 nor band 4. The small number of band 3 and 4 sources is due only to the limited coverage of the band 3 and 4 images. 

\subsection{MIPS 24~\micron\ Photometry}\label{subsec:MIPSphot}
We performed photometry on the MIPS 24~\micron\ images using SExtractor, finding 1882 sources. Parameters and details of the process are given in Appendix \ref{appendix:ExtractPhotoIRAC}.

\subsection{MMIRS Photometry} \label{subsec:MMIRSphot}
We performed photometry on the MMIRS images using SExtractor with parameters optimized for the spatial resolution and background characteristics of the MMIRS images. The details of the process are given in Appendix \ref{appendix:ExtractPhotoMMIRS}.
After calibration, all sources in each band were merged into one catalog using a cross matching distance of 2\arcsec. We obtained catalogs with 158,526 J-band sources, 177,510 H-band sources, and 7,613 K-band sources.

\subsection{Cross Matching} \label{subsec:crossmatch}

Using our complete IRAC catalog obtained via the methods described in Section~\ref{subsec:phot}, we cross matched our IRAC sources with the \textit{WISE} and 2MASS databases described in Sections~\ref{subsec:WISE} and \ref{subsec:2MASS}. We used a cross matching distance of 1\farcs5 and found 13\% of our sources had corresponding \textit{WISE} and 2MASS photometry, 6\% of our sources had 2MASS but no \textit{WISE} photometry, 4\% of our sources had \textit{WISE} but no 2MASS photometry, and 77\% had neither \textit{WISE} nor 2MASS photometry. While we did not investigate the false matching rate rigorously, we do not expect it to be significant considering that the mean positional error for IRAC sources with respect to 2MASS sources is 0\farcs25 \citep{Fazio2004}.

We also cross matched our catalog with the extracted catalog of MIPS 24~\micron\ photometry described in Section~\ref{subsec:phot} using a radius of 2\arcsec. We found 485 MIPS sources with corresponding IRAC photometry. This small number of matched sources is again due to the limited coverage of the MIPS images. 

Lastly, we cross matched the MMIRS near-IR photometry with the compiled catalog. Matching MMIRS required additional consideration to account for the difference in MMIRS and IRAC resolution. We used a matching distance of 1\farcs5 and flagged any IRAC sources with multiple close MMIRS source – a potential indication of unresolved stars in the IRAC images. If one of the close MMIRS sources was at least one magnitude brighter than the other neighbors, we assumed the IRAC photometry was not detrimentally affected by the unresolved neighbors and we flagged the source as a good multi-match. If all of the MMIRS close neighbors were of comparable brightness, then the IRAC photometry likely did not accurately correspond to the MMIRS photometry and we flagged such sources as bad multi-matches and did not include them in the MMIRS YSO selection in Section~\ref{subsec:MMIRS+IRACysosel}. Of the 146,344 MMIRS source that matched to an IRAC source 4\% had good multi-matching and 5\% had bad multi-matching.

\section{YSO Identification and Classification}\label{sec:YSO}
We identified the YSO candidates in our compiled catalog by selecting objects with excess IR emission, a result of the reprocessed stellar radiation from the cool, dusty material in the surrounding circumstellar disks and envelopes. After removing background sources, we selected the YSO candidates using various color-color cuts to distinguish them from cool or reddened stars \citep{Allen2004, Gutermuth2004, Winston2007, Winston2020}. Hereafter  we  will  refer  to  the  YSO  candidates as ``YSOs,” but a definitive classification would require a more detailed analysis of the spectra and other characteristics of each object. 

In this section, we present the criteria used in the YSO selections for each subset of our catalog. After identifying the YSOs, we classify their evolutionary stages based on the slopes of their SEDs between 2 and 20~\micron. 
\subsection{2MASS+IRAC} \label{subsec:2MASS+IRACysosel}
\subsubsection{Background Sources} \label{subsec:2MASScontaminants}
 We anticipated that some of the detected sources with excess IR emission would be active galactic nuclei (AGN) or star forming galaxies (SFG) with polycyclic aromatic hydrocarbon (PAH) emission. Foreground PAH emission is also a potential source of aperture contamination that could affect a source's measured photometry. The region will also have many main sequence stars in the foreground and background that are unrelated to the Cep~OB4 clusters.
We refer to all such sources as background sources (acknowledging that some stars and PAH emission may actually be foreground contamination) and removed them before proceeding with the YSO selection. We follow the selection criteria described in \citet{Winston2020}, which is adapted from \citet{Winston2019, Gutermuth2008b, Gutermuth2009} and reported in Appendix~\ref{appendix:2MASSIRAC}.

\begin{figure}
\begin{center}
\includegraphics[width=\linewidth,angle=0]{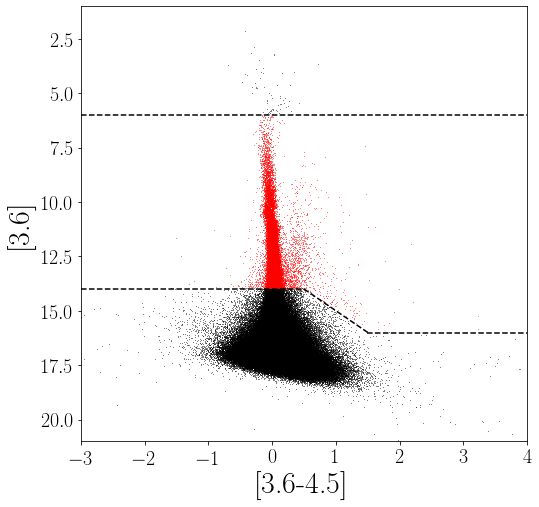}
\caption{The results of the IRAC [3.6] versus [3.6-4.5] identification of background sources. The dashed black lines represent the selection criteria presented in Section~\ref{subsec:2MASScontaminants} and Appendix~\ref{appendix:2MASSIRAC}. The background sources eliminated from the YSO selection are shown in grey and the sources considered in the YSO evaluation are shown in red. \label{fig:contaminants}}
\end{center}
\end{figure}

The result of the background source identification is shown in Figure~\ref{fig:contaminants}. Of the 752,615 sources, we classified 711,152 (94.5\%) as background sources and 41,463 (5.5\%) as potential YSOs to be evaluated further. In the following sections, the 2MASS+IRAC YSO selections were based only on these 41,463 potential YSOs. 

\subsubsection{2MASS+IRAC YSO Selection} \label{subsec:IRACYSO}
We selected YSOs based on excess IR emission using 2MASS+IRAC bands in a variety of color-color combinations. Of the 41,463 sources evaluated, 99\% had 2MASS photometry. We calculated the extinction coefficient $A_{K_s}$ for each of these sources following the IR extinction laws described in \citet{Flaherty2007}. We removed the 29 sources with non-valid values of extinction from our YSO selection, leaving 41,030 non-contaminant sources with good extinction values and 2MASS photometry.

\begin{figure*}
\begin{center}
\includegraphics[width=\linewidth,angle=0]{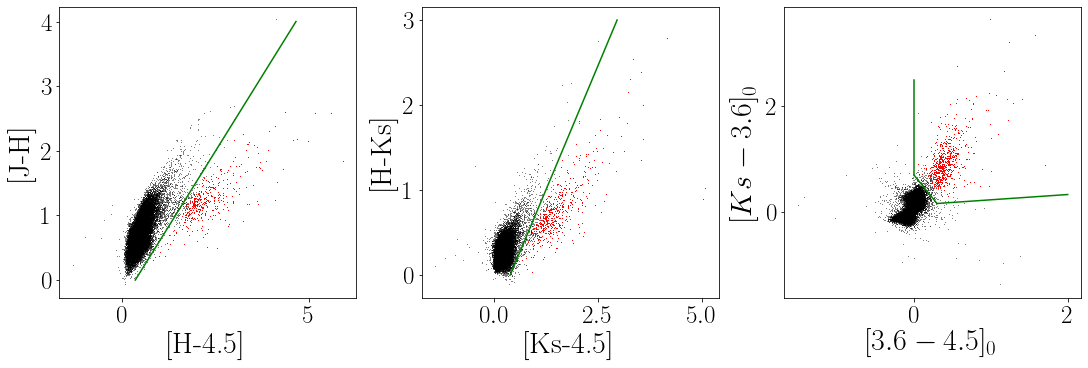}
\caption{The results of the 2MASS+IRAC YSO selection. The green lines show the selection criteria described in Appendix~\ref{appendix:2MASSIRAC}. The field sources not identified as YSOs are plotted in grey and the identified YSOs are plotted in red. The YSOs are identified by their excess IR emission. We identified 388 YSOs in the leftmost selection, 514 YSOs in the middle selection, and 656 YSOs in the rightmost selection. \label{fig:2MASSyso}}
\end{center}
\end{figure*}

We adapted the selection criteria from \citet{Winston2020}. We aimed to include the maximal number of YSOs and so made the cutoff less conservative to encompass more sources. However, we wanted to minimize the number of non-YSO field stars selected, and so we examined the spatial distribution of the selected sources to confirm that the additional sources were not scattered randomly about the field, but instead were concentrated in clusters or located near pillars. This spatial distribution provided an additional confirmation of the validity of our selection. The criteria are reported in Appendix~\ref{appendix:2MASSIRAC} and the results are shown in Figure~\ref{fig:2MASSyso}. Of the 41,030 candidates, 719 were selected as YSOs, about 1.7\%. 

\subsection{WISE} \label{subsec:WISEysosel}
\subsubsection{Background Sources} \label{subsec:WISEcontaminants}
We followed the \textit{WISE} contaminant removal procedure described in \citet{Winston2020} based on \citet{Fischer2016} and \citet{Koenig2014} to identify AGN and SFG in the subset of our catalog with \textit{WISE} photometry. The specific criteria for selection are described in Appendix~\ref{appendix:WISE}. Of the 126,835 sources with \textit{WISE} photometry, 116,796 were identified as AGN and 68,634 were identified as SFG. This left 6,526 remaining \textit{WISE} YSO candidates.

\subsubsection{WISE YSO Selection} \label{subsec:WISEYSO}
We identified YSOs from the remaining \textit{WISE} YSO candidates using the four \textit{WISE} photometric bands. The selection cuts are reported in Appendix~\ref{appendix:WISE} and shown in Figure~\ref{fig:WISE}. Of the 6,526 candidates, we identified 235 \textit{WISE} YSOs, about 3.6\%.

\begin{figure*}
\begin{center}
\includegraphics[width=\linewidth,angle=0]{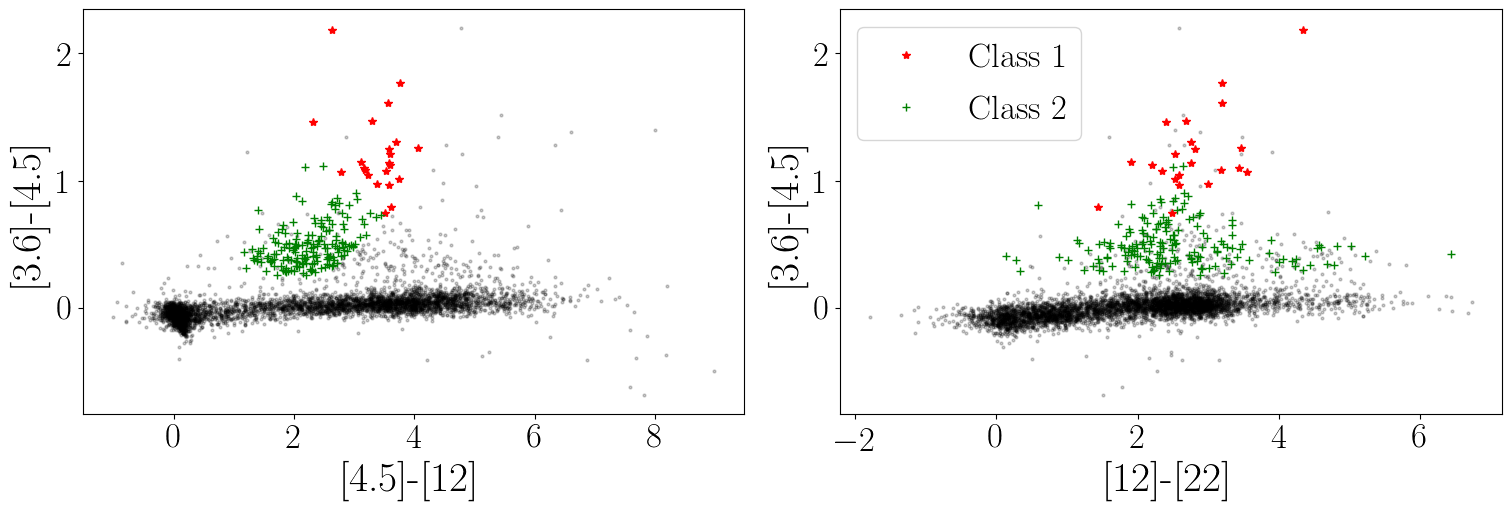}
\caption{The results of the \textit{WISE} \& IRAC YSO selection described in Appendix~\ref{appendix:WISEIRAC} and Section~\ref{subsec:WISE+IRACysosel}. The red points are YSOs identified as Class 1, the green points are identified as Class 2, and the black points are the non-AGN non-SFG non-YSO sources. We identified 22 Class 1 sources and 174 Class 2 sources.  \label{fig:WISE}}
\end{center}
\end{figure*}

\subsection{WISE \& IRAC} \label{subsec:WISE+IRACysosel}
We repeated the procedure for AGN and SFG removal and YSO selection described in Section~\ref{subsec:WISEysosel} using the IRAC photometry for 3.6 and 4.5~\micron\ instead of the \textit{WISE} bands 1 \& 2. This additional selection uses the shorter wavelength IRAC magnitudes, which may not have been reliably detected with \textit{WISE}. We also excluded objects identified as background sources using IRAC and 2MASS in Section~\ref{subsec:2MASS+IRACysosel} from the YSO selection in this section.

Using these photometry, we found that of the 126,835 sources with \textit{WISE} photometry, 111,879 were identified as AGN and 59,491 were identified as SFG. An additional 5,625 that were not identified as AGN or SFG were identified as background sources in Section~\ref{subsec:2MASS+IRACysosel} and were removed. This left 5,217 YSO candidates. 
Applying the same selection criteria as in Section~\ref{subsec:WISE+IRACysosel}, we identified 196 YSOs, about 3.7\%.

\subsection{MMIRS \& IRAC} \label{subsec:MMIRS+IRACysosel}
We repeated the background source removal and YSO selection described in Section~\ref{subsec:2MASS+IRACysosel}, using MMIRS photometry in place of 2MASS. We only included sources with MMIRS photometry below the saturation limit: 12.7 magnitudes for H-band, 12.5 magnitudes for J-band, and 11.1 magnitudes for K-band. We also excluded sources that had corresponding 2MASS photometry above these limits as such sources were saturated and thus had unreliable photometry in the MMIRS images. These restrictions left 6,803 non-background sources with MMIRS photometry in the H-band, 8,501 J-band sources, and 1,408 sources in K-band. Of the 10,230 sources with good MMIRS photometry in at least one band, 232 were selected as YSOs, about 2.3\%. 

\subsubsection{Comparing MMIRS and 2MASS YSOs}\label{subsubsec:MMIRS+2MASS}

The MMIRS YSO selection yielded 53 additional YSOs that were not identified using 2MASS+IRAC in Section~\ref{subsec:2MASS+IRACysosel}, an expected result as the MMIRS photometry is generally more reliable due to its higher resolution. There were 28 YSOs identified using 2MASS+IRAC that had good MMIRS photometry in all of the necessary bands but were not selected as YSOs in the MMIRS cuts. As this represented a relatively small portion of the YSOs and there was no discernible trend in MMIRS and 2MASS magnitude differences in these sources, we attributed this difference to potential photometric uncertainties and intrinsic variability in the stars. We flagged these YSOs in our catalog, but retained them for subsequent analysis. 

Another useful aspect of the MMIRS photometry is that it can reveal the presence of unresolved 2MASS sources that were selected as YSOs. Of the 221 2MASS YSOs with good corresponding MMIRS photometry in at least one band used for 2MASS selection that were not selected as MMIRS YSOs, 6 were flagged as MMIRS sources with good multi-matching and 11 (with one overlap) were flagged as MMIRS sources with bad multi-matching as defined in Section~\ref{subsec:crossmatch}. This represents a total of 16 2MASS YSOs that are potentially multiple unresolved stars. We flagged the sources as such, but retain them in our YSO catalog for subsequent steps of the analysis. 

\subsection{IRAC 5.8 \& 8.0~\micron}\label{subsec:IRAC34}
In the regions with coverage in IRAC bands 3 and 4, we selected YSOs following the methods of \citet{Winston2019}. The selection criteria are described in Appendix~\ref{appendix:IRAC34}. We identified 60 total YSOs, 9 of which were not identified using the selections in previous sections.

\subsection{MIPS 24 \micron}\label{subsec:YSOmips}
In the regions with MIPS 24~\micron\ coverage, we selected YSOs following the methods of \citet{Winston2019}. The selection criteria are described in Appendix~\ref{appendix:MIPS}. We identified 39 total YSOs, 10 of which were not identified using IRAC bands 1 and 2, 2MASS, MMIRS, and \textit{WISE} photometry. 
%We note that while our selections actually identified 39 YSOs, we removed two sources that visually appeared to be false detections obtained in the source extraction process described in Section~\ref{subsec:phot}.

\renewcommand{\arraystretch}{0.9}
\begin{deluxetable}{ccl}
%\tabletypesize{\scriptsize}
\tabletypesize{\footnotesize}
\tablecolumns{3}
\tablecaption{Photometry Table Description\label{tab:phot}}
\tablehead{\colhead{Column}\\[-0.3cm]  
\colhead{Number} & \colhead{Column ID}  & \colhead{Description}  
}
\startdata
0   &  Name &  Source Name      \\  
1   &  RAdeg            &   Right ascension 2000 (degrees)   \\  
2   &  DEdeg           &  Declination 2000 (degrees)    \\  
3   &  MAGI1        &  IRAC 3.6 \micron\ magnitude \\
4   &  e\_MAGI1    &  IRAC 3.6 \micron\ uncertainty \\
5   &  MAGI2       &  IRAC 4.5 \micron\ magnitude \\
6   &  e\_MAGI2     &  IRAC 4.5 \micron\ uncertainty \\
7   &  MAGI3        &  IRAC 5.8 \micron\ magnitude \\
8   &  e\_MAGI3    &  IRAC 5.8 \micron\ uncertainty \\
9   &  MAGI4        &  IRAC 8.0 \micron\ magnitude \\
10  &  e\_MAGI4    &  IRAC 8.0 \micron\ uncertainty \\
11  &  jm2MASS     & 2MASS J-band magnitude   \\  
12  &  e\_jm2MASS  &   2MASS J-band uncertainty     \\  
13  &  hm2MASS     &  2MASS H-band magnitude    \\  
14  &  e\_hm2MASS  &  2MASS H-band uncertainty     \\  
15  &  km2MASS     &  2MASS K-band magnitude    \\  
16  &  e\_km2MASS   &  2MASS K-band uncertainty     \\  
17  &  f\_jm2MASS  &  2MASS J-band flag     \\  
18  &  f\_hm2MASS  &  2MASS H-band flag     \\  
19  &  f\_km2MASS  &  2MASS K-band flag     \\  
20  &  w1mpro        & WISE band 1 magnitude     \\  
21  &  e\_w1mpro     & WISE band 1 uncertainty     \\  
22  &  w2mpro        & WISE band 2 magnitude    \\  
23  &  e\_w2mpro     & WISE band 2 uncertainty     \\  
24  &  w3mpro        & WISE band 3 magnitude    \\  
25  &  e\_w3mpro     & WISE band 3 uncertainty     \\  
26  &  w4mpro        & WISE band 4 magnitude    \\  
27  &  e\_w4mpro     & WISE band 4 uncertainty     \\  
28  &  f\_w1mpro  &  WISE band 1 flag     \\  
29  &  f\_w2mpro  &  WISE band 2 flag     \\  
30  &  f\_w3mpro  &  WISE band 3 flag     \\  
31  &  f\_w4mpro  &  WISE band 4 flag     \\  
32  &  MAGMIPS      & MIPS 24 \micron\ magnitude    \\
33  &  e\_MAGMIPS  & MIPS 24 \micron\ uncertainty   \\  
34  &  MAGJMMIRS        & MMIRS J-band magnitude   \\  
35  &  e\_MAGJMMIRS   & MMIRS J-band uncertainty   \\  
36  &  MAGHMMIRS       & MMIRS H-band magnitude   \\  
37  &  e\_MAGHMMIRS   & MMIRS H-band uncertainty   \\  
38  &  MAGKMMIRS        & MMIRS K-band magnitude   \\  
39  &  e\_MAGKMMIRS   & MMIRS K-band uncertainty   \\  
40  &  YSO   & YSO flag   \\  
\enddata
\end{deluxetable}

\subsection{Combined YSO Catalog}\label{subsec:YSOcombcat}
\begin{figure}
\begin{center}
\includegraphics[width=\linewidth,angle=0]{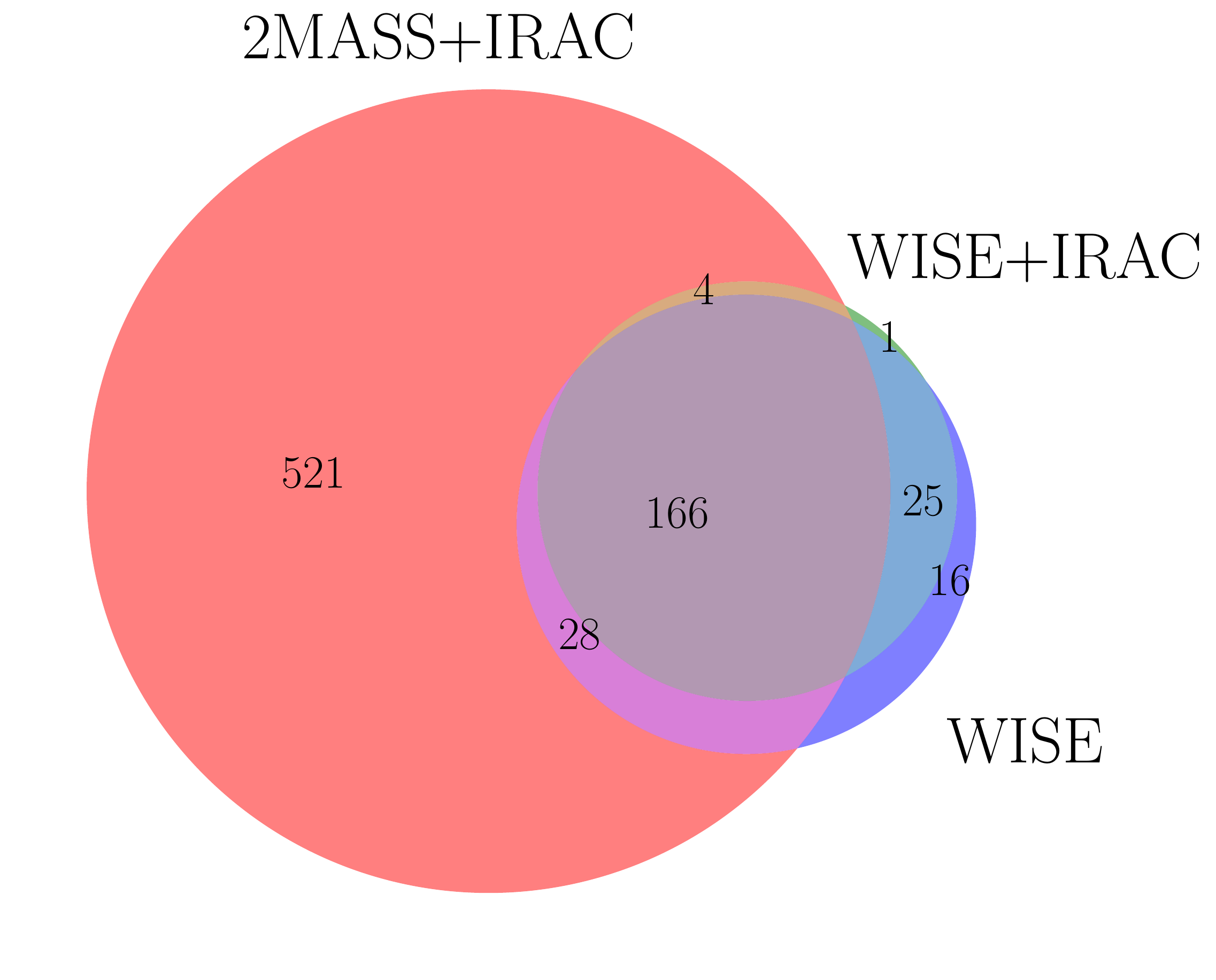}
\caption{A Venn diagram showing the breakdown of YSO classification methods (excluding the selections based on IRAC bands 3 and 4 described in Section~\ref{subsec:IRAC34}, MIPS 24~\micron\ described in Section~\ref{subsec:YSOmips}, and MMIRS described in Section~\ref{subsec:MMIRS+IRACysosel}, which are all restricted by limited spatial coverage). The vast majority of the YSOs were identified using 2MASS+IRAC photometry only. \label{fig:piechart}}
\end{center}
\end{figure}

After performing each of the selections described above, we merged all of the YSOs into a single catalog. Five YSOs were removed upon visual inspection as they appeared to have unreliable photometry resulting from poor extraction in especially nebulous regions. We describe the resulting photometric catalog in Table~\ref{tab:phot}. Our final YSO catalog included 821 sources. 22.0\% identified in all three selections, 68.1\% identified in 2MASS+IRAC only, 2.1\% in \textit{WISE} only, 0.1\% in WISE+IRAC only. A breakdown of the selection results are shown in Figure~\ref{fig:piechart}. Note that this figure does not include the results of the YSO selections using IRAC bands 3 and 4, MIPS 24~\micron, or MMIRS. We excluded these sources from the summary chart as their selection statistics are representative only of the limited field coverage.

In following sections, when we reference our sample of YSOs, we refer to this merged catalog.

\begin{figure}
\begin{center}
\includegraphics[width=\linewidth,angle=0]{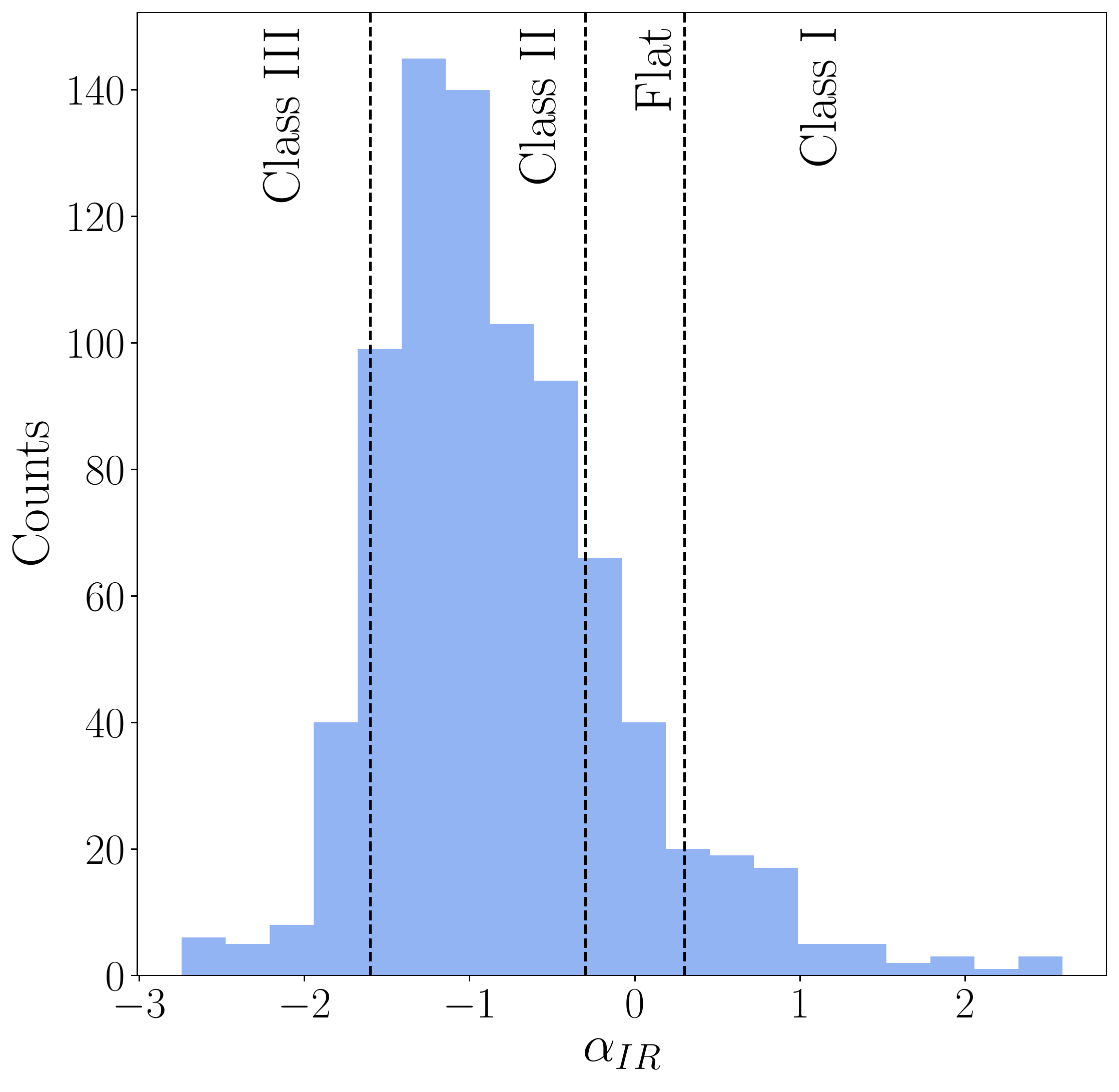}
\caption{A histogram of the SED slopes $\alpha_{IR}$ of the YSOs. The black dashed lines at $\alpha_{IR}=-1,6, -0.3, 0.3$ show the cutoffs used in the evolutionary classification. Class 0 and I objects have $\alpha_{IR}\geq0.3$, flat spectrum sources have $-0.3<\alpha_{IR}<0.3$, Class II sources have $-1.6<\alpha_{IR}\leq-0.3$, and Class III have $\alpha_{IR}\leq-1.6$. \label{fig:alphahist}}
\end{center}
\end{figure}

\subsubsection{Evolutionary Classification} \label{subsec:evoclass}
YSOs are typically classified into a number of evolutionary stages, ranging from embedded protostars to pre-main sequence YSOs \citep{McKee2007}. We classified the YSOs in our sample following the classification scheme developed by \citet{Andre2000}.

We calculated the slope of the SED over the wavelength range of 2-20~\micron\ defined as $\alpha_{IR}\equiv \frac{d\log{F_\lambda}}{d\log{\lambda}}$ \citep{Lada1987}. This included photometry in the K-band from 2MASS or MMIRS, in bands 3.6, 4.5, 5.4, and 8.0 \micron\ from IRAC, and in bands 12 and 22 \micron\ from \textit{WISE}.
% For the journal paper, we don't need to explain the flux calculation
% To calculate $\alpha_{IR}$, we converted our measured magnitudes to $F_\nu$ following the equation:
% $$F_\nu = F_{\nu_0}\cdot 10^{-m/2.5}$$
% where $F_{\nu_0}$ is the zero-magnitude flux for a given band (see Table~\ref{tab:fnu} for the values of $F_{\nu_0}$). We convert from $F_\nu$ to $F_\lambda$ using the relation:
% $$\lambda F_\lambda=\nu \cdot F_\nu$$
% After calculating these conversions, we 
We fit a line to the log-scaled SEDs using the mean squared error minimization algorithm implemented in NumPy's \texttt{polyfit}. To account for potential contamination in the lower resolution \textit{WISE} photometry, we identified sources whose $\alpha_{IR}$ changed signs when including WISE 12 and 22~\micron\ photometry. We fit $\alpha_{IR}$ on these 115 sources using only K-band and IRAC photometry. We used MMIRS photometry in place of 2MASS photometry for the K-band when available. 

\begin{figure*}
\begin{center}
\includegraphics[width=0.75\linewidth,angle=0]{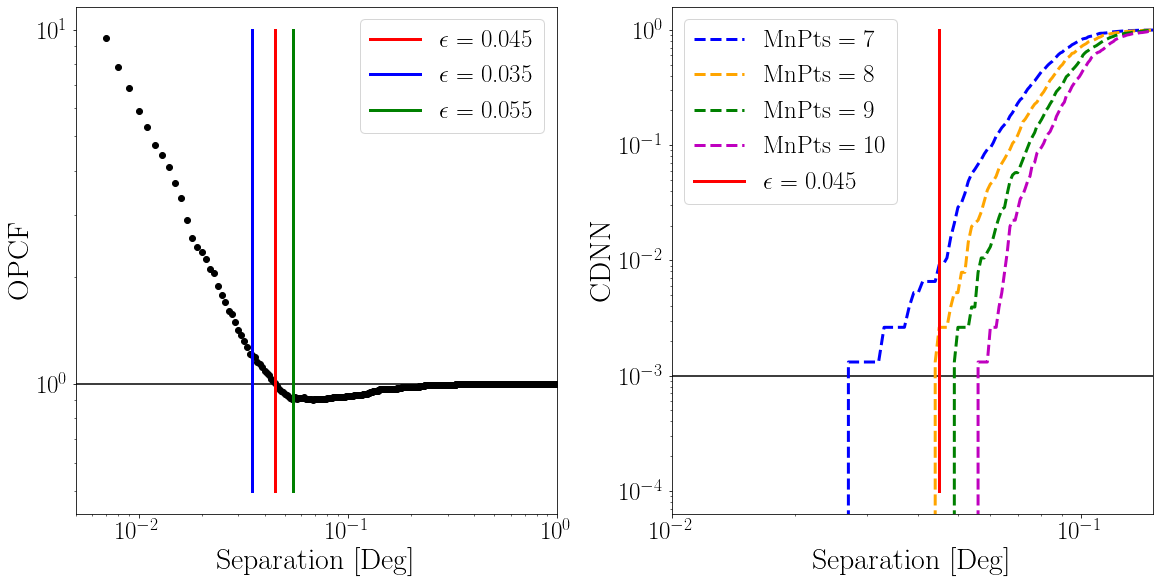}
\caption{(Left) The one point correlation function used to determine the optimal value of $\epsilon$, the bandwidth used for DBSCAN. The red line indicates the chosen value of $\epsilon = 0.045$ selected at the point with OPCF $\approx$ 1. (Right) The cumulative distribution of nearest neighbors used to determined the optimal value of MnPts. When MnPts = 9, the probability of finding a random cluster for $\epsilon=0.045$ is less than 0.1\%. \label{fig:heuristics}}
\end{center}
\end{figure*}

We used the $\alpha_{IR}$ cutoffs given in \citet{Andre2000} to classify the YSOs:  Class 0 and I objects (protostellar) have a rising slope, $\alpha_{IR}\geq0.3$, Class II sources have decreasing slopes between $-1.6<\alpha_{IR}\leq-0.3$, and Class III have $\alpha_{IR}\leq-1.6$. We classify sources with $-0.3<\alpha_{IR}<0.3$ as flat spectrum sources.

We classified 67 of our YSOs as Class I, 103 as flat spectrum sources, 569 as Class II, and 82 as Class III. A histogram of these results is shown in Figure~\ref{fig:alphahist}.

\begin{deluxetable*}{lrcccrcc}
\tablecaption{Cep~OB4 YSOs\label{tab:YSO}}
\tablewidth{0pt}
\tablehead{ 
\colhead{YSO ID} & \colhead{R.A.} & 
\colhead{Decl} & 
\colhead{YSO Selection Method\tablenotemark{a}}   & 
\colhead{Evo. Class} & 
\colhead{$\alpha_{IR}$}  &
\colhead{$\sigma_{\alpha_{IR}}$}  &
\colhead{Cluster Membership} \\[-0.3cm]
\colhead{}  & \colhead{deg}  & 
\colhead{deg} & 
\colhead{}  & 
\colhead{} &  
\colhead{} &
\colhead{} &
\colhead{} 
}
\startdata
SSTCOB4 J23470610+6745164 	 & 356.7754205 	 & 67.7545648 	 & 000100 	 & II 	 & -0.91 	 & 0.37 	 & ---\\
SSTCOB4 J23473479+6751001 	 & 356.8949644 	 & 67.8500461 	 & 000100 	 & Flat  & -0.02 	 & 0.12 	 & ---\\
SSTCOB4 J00095102+6842133 	 & 2.4625862 	 & 68.7037036 	 & 000100 	 & Flat  & -0.13 	 & 0.30 	 & ---\\
SSTCOB4 J00102105+6845167 	 & 2.5877272 	 & 68.7546457 	 & 000100 	 & III 	 & -2.03 	 & 0.17 	 & ---\\
SSTCOB4 J00114782+6847195 	 & 2.9492560 	 & 68.7887726 	 & 100100 	 & Flat  & -0.25 	 & 0.05 	 & ---\\
SSTCOB4 J00115485+6806090 	 & 2.9785559 	 & 68.1025245 	 & 000100 	 & I 	 & 0.66 	 & 0.14 	 & ---\\
SSTCOB4 J23474717+6748538 	 & 356.9465785 	 & 67.8149459 	 & 000100 	 & II 	 & -1.33 	 & 0.36 	 & ---\\
SSTCOB4 J23484602+6803394 	 & 357.1917535 	 & 68.0609588 	 & 000100 	 & I 	 & 0.85 	 & 0.18 	 & ---\\
SSTCOB4 J23465418+6817350 	 & 356.7257701 	 & 68.2930714 	 & 000100 	 & I 	 & 0.48 	 & 0.20 	 & ---\\
SSTCOB4 J23473180+6818507 	 & 356.8825085 	 & 68.3140956 	 & 100000 	 & III 	 & -2.25 	 & 0.47 	 & ---\\
\enddata
\tablenotetext{}{(This table is available in its entirety in machine-readable form.)}
\tablenotetext{a}{The binary flag describes the selection method(s) used to identify the YSO with digits representing 2MASS, \textit{WISE}, IRAC+\textit{WISE}, MMIRS, IRAC bands 3 and 4, and MIPS selections respectively.}
\end{deluxetable*}

\begin{figure*}
\begin{center}
\includegraphics[width=\linewidth,angle=0]{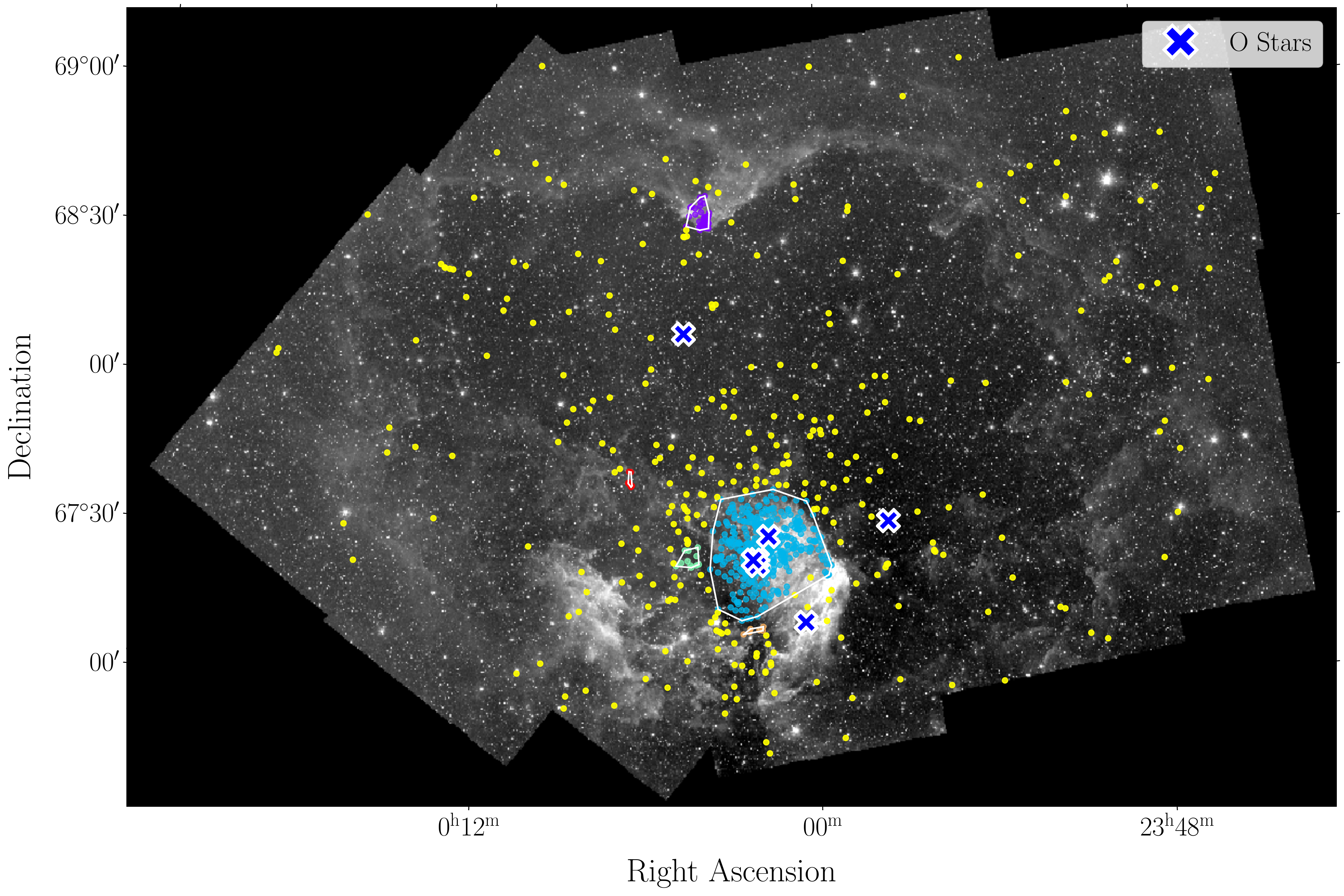}
\caption{The identified YSOs and clusters shown over the IRAC 3.6~\micron\ image. The five clusters are plotted in purple, blue, green, orange and red respectively. The YSOs that are not members of any cluster are shown in yellow. The convex hulls for each cluster are overlayed in white. The largest cluster (blue) is located in the Be~59 region. The characteristics of the clusters are reported in Table~\ref{tab:clusters}. The location of O-type stars cataloged in \citet{Skiff2014} are plotted as blue x's outlined in white.\label{fig:clusters}}
\end{center}
\end{figure*}

\begin{figure}
\begin{center}
\includegraphics[width=0.85\linewidth,angle=0]{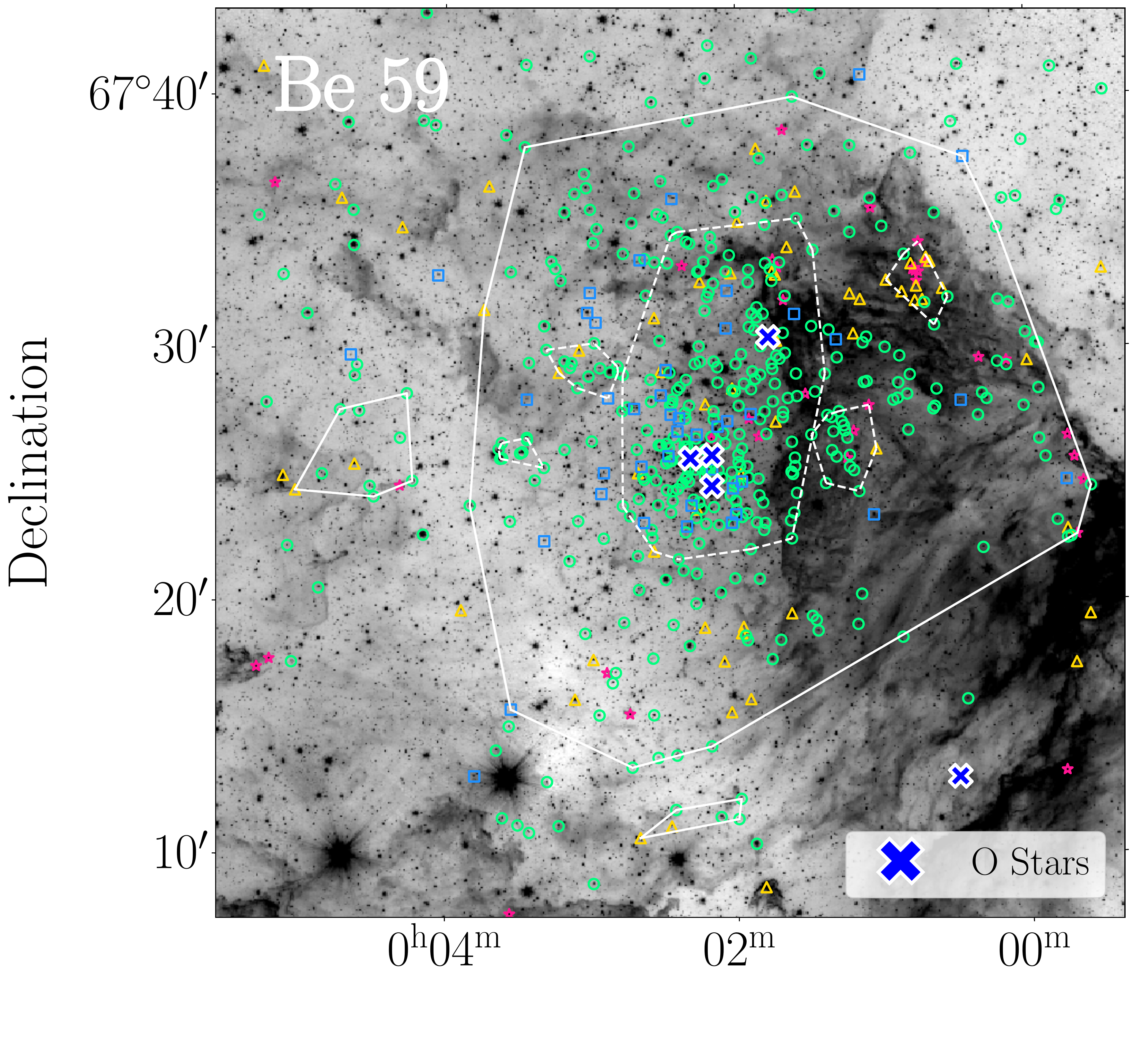}
\includegraphics[width=0.85\linewidth,angle=0]{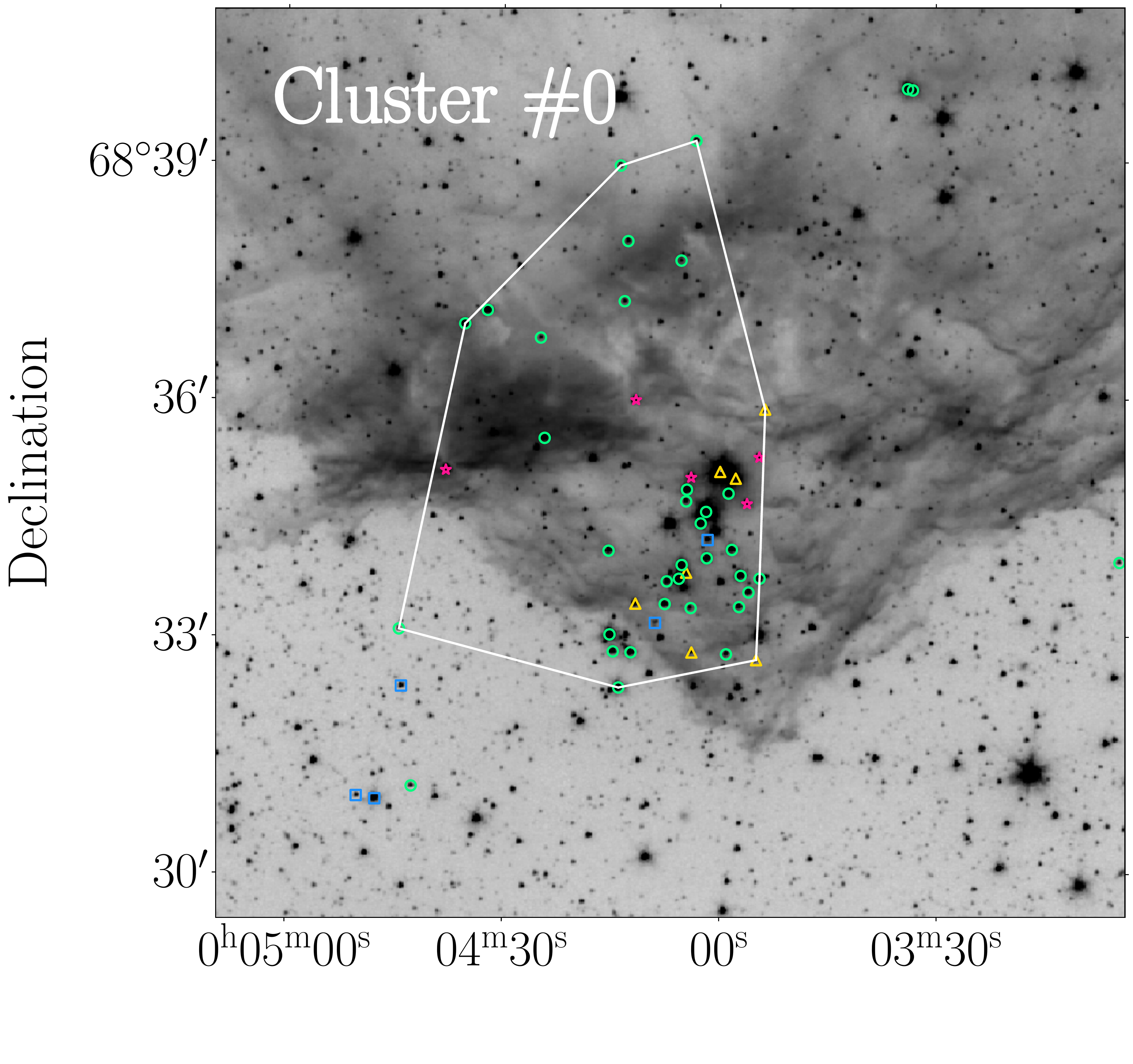}
\includegraphics[width=0.85\linewidth,angle=0]{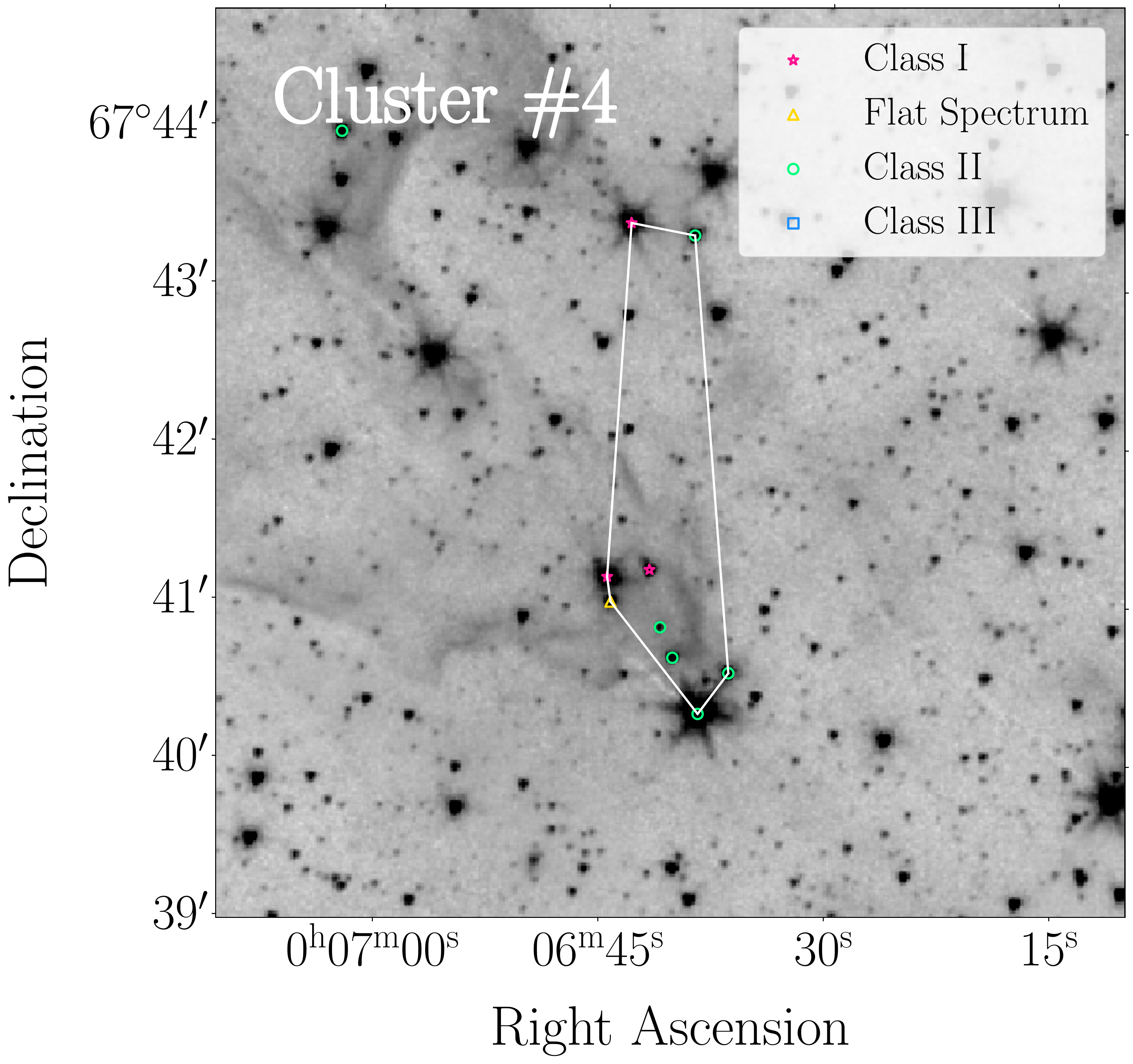}
\caption{The five identified YSO clusters plotted over an inverted gray scale image at 3.6~\micron. The color and shapes of the YSOs indicate the evolutionary class as determined in Section~\ref{subsec:evoclass}. The white lines show the convex hulls of each cluster, with all cluster members included in the indicated region. The upper plot shows the Be~59 region with the convex hulls of the subclusters identified in Section~\ref{subsubsec:subcluster} plotted in white dashed lines and the same O stars as in Figure~\ref{fig:clusters}. \label{fig:clusterstogether}}
\end{center}
\end{figure}

%\vspace{-3em}
\section{YSO Distribution}\label{sec:distribution}
\subsection{Clustering}\label{subsec:clustering}
\subsubsection{DBSCAN}\label{subsubsec:dbscan}

We expected the identified YSOs to be clustered together (in the area of Be\ 59 for example) or in molecular pillars towards the edges of the \ion{H}{2} region and not dispersed randomly over the field. To analyse this clustering, we used a method called ``Density-based spatial clustering of applications with noise” \citep[DBSCAN;][]{Xu1997}. This method identifies clusters by grouping points with many neighbors and flagging points with few neighbors as outliers. We used the DBSCAN implementation in Python's SKLEARN package. 

The DBSCAN algorithm requires two input parameters: $\epsilon$ and MnPts. The scaling size parameter, $\epsilon$ controls the bandwidth used to classify close neighbors. MnPts defines the smallest number of samples in a neighborhood required for a point to be considered as a core point. To determine the optimal values of these parameters, we followed the approach of \citet{Winston2019, Winston2020} based on the analysis of the Taurus region done by \citet{Joncour2018}.

We chose $\epsilon$ as the turnoff value in the one-point correlation function (OPCF). The OPCF is calculated as the ratio of the cumulative distribution of nearest neighbors of the true distribution and a randomly generated distribution of the same number of sources spread over the same area. For a given separation $\epsilon$, the cumulative distribution of nearest neighbors counts the number of sources with a neighbor within a radius of $\epsilon$. When the OPCF is greater than 1, there are significantly more nearest neighbor pairings than there are in the randomly generated sample. When the OPCF is less than or equal to 1, the number of nearest neighbors in the true distribution is comparable to that of the random distribution. Therefore, we choose the value of $\epsilon =$ 0\fdg045 \ at the turnoff point in the OPCF as seen in Figure~\ref{fig:heuristics} to minimize detections of random clustering. 

The value of MnPts was determined by choosing the smallest number of points that results in a value of 0.001 for the cumulative distribution of nearest neighbors evaluated at $\epsilon=$ 0\fdg045. This represents a probability of 0.001 of randomly finding a cluster of size MnPts for a scale factor of $\epsilon$. We chose MnPts = 9 as shown in Figure~\ref{fig:heuristics}. 

While the spatial distribution of YSOs identified using MMIRS, MIPS, and IRAC bands 3 and 4 is biased by the limited coverage of these images, we did not find a significant difference in clustering results when they were included. This is likely due to the concentration of coverage in Be~59 and the northern region around IRAS00013+6817 where there are dense clusters of 2MASS and \textit{WISE} YSOs. For this reason, we did not exclude these sources from our clustering analysis.

\begin{figure}
\begin{center}
\includegraphics[width=\linewidth,angle=0]{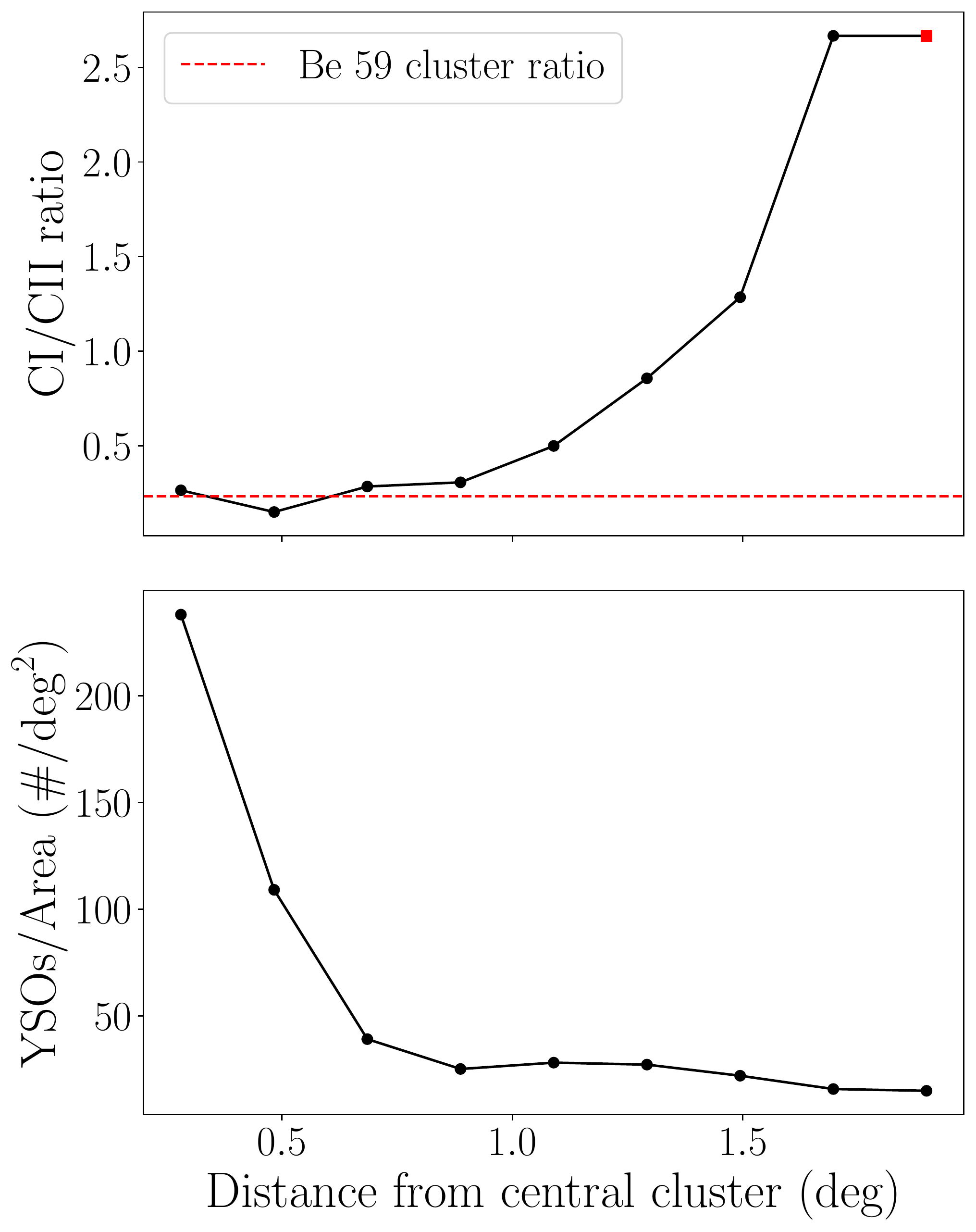}
\caption{The distribution of unclustered YSOs. The top panel shows the Class~I/Class~II (CI/CII) ratio of YSOs in annular bins as a function of projected distance from the center of the Be~59 cluster found in Section~\ref{subsec:clustering}. The red horizontal line indicates the CI/CII ratio of sources in the three central clusters. The value in the rightmost bin – marked by a red square – is undefined as there were no Class~II sources in the final bin, and it is set to the same value as the previous bin. The bottom panel shows the density of the unclustered YSOs, normalized by the area in each annulus covered by the IRAC 3.6 \micron\ mosaic.  Both of these distributions provide evidence that many of the unclustered YSOs in the region's center may have formed in the central clusters, and there is an age gradient with a higher fraction of younger YSOs at greater distances from Be~59, as described in Section~\ref{subsec:unclusteredysos}. \label{fig:unclustered}}
\end{center}
\end{figure}

\begin{figure*}
\begin{center}
\includegraphics[width=0.48\linewidth,angle=0]{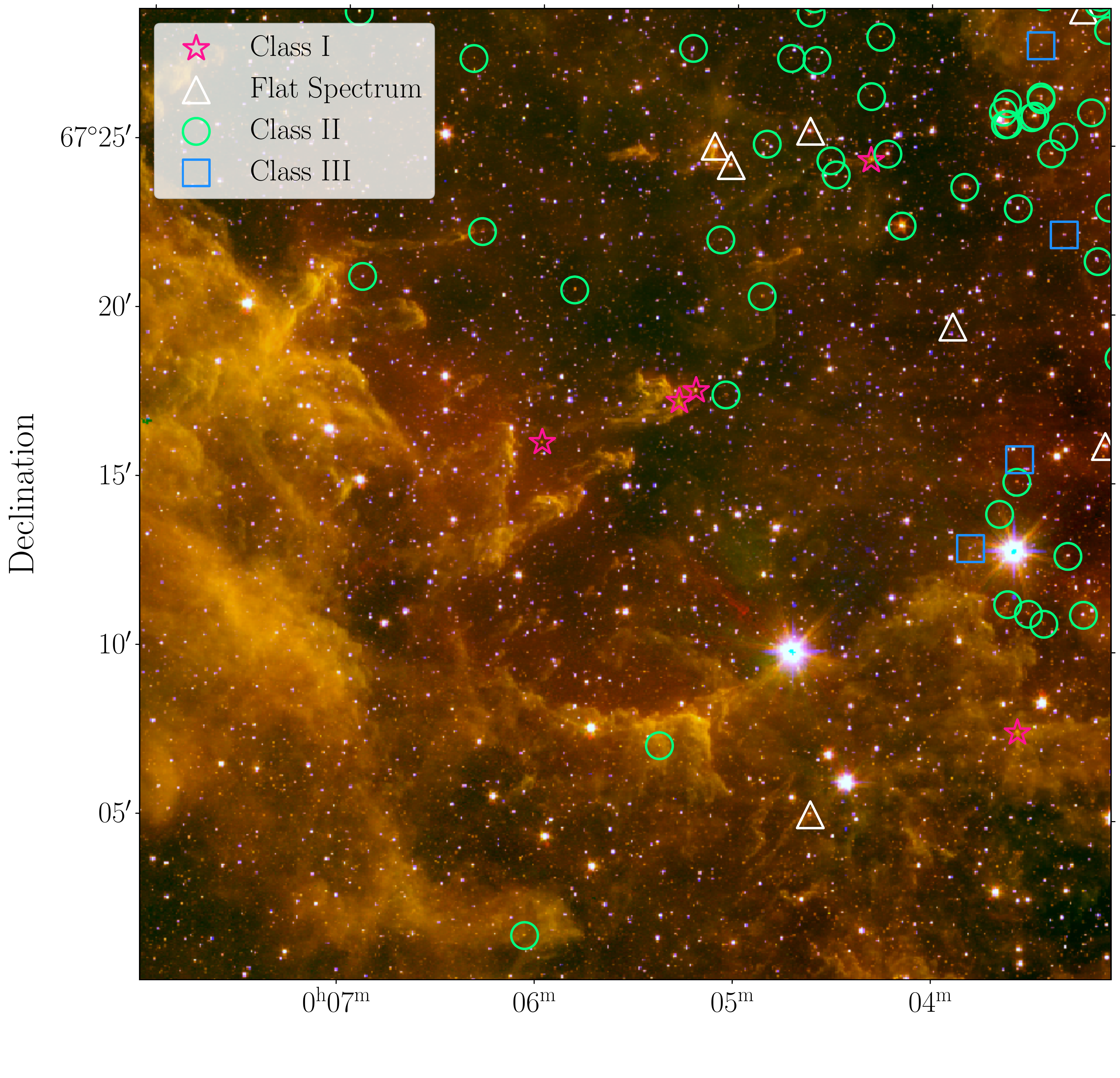}
\includegraphics[width=0.48\linewidth,angle=0]{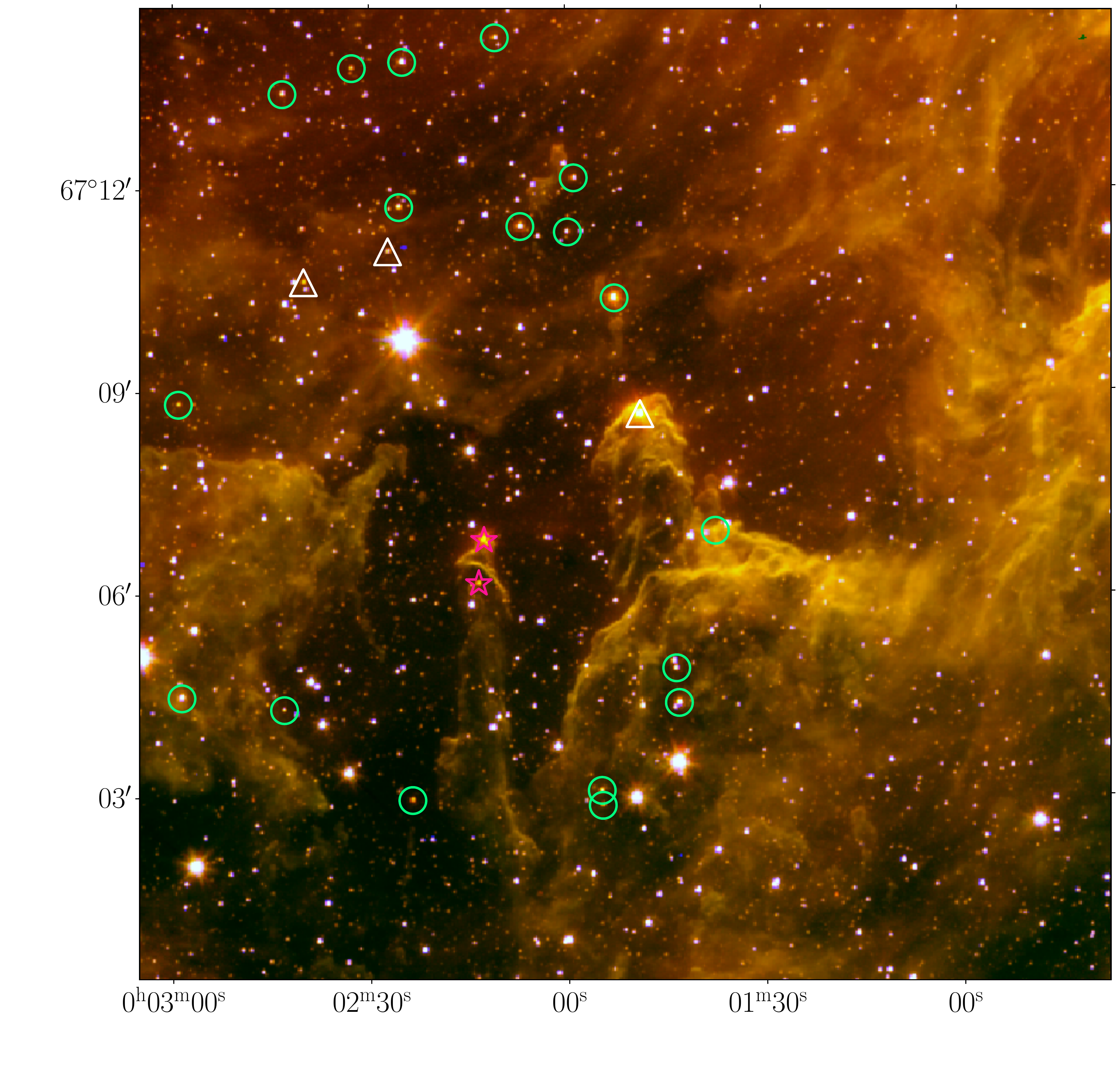}
\includegraphics[width=0.48\linewidth,angle=0]{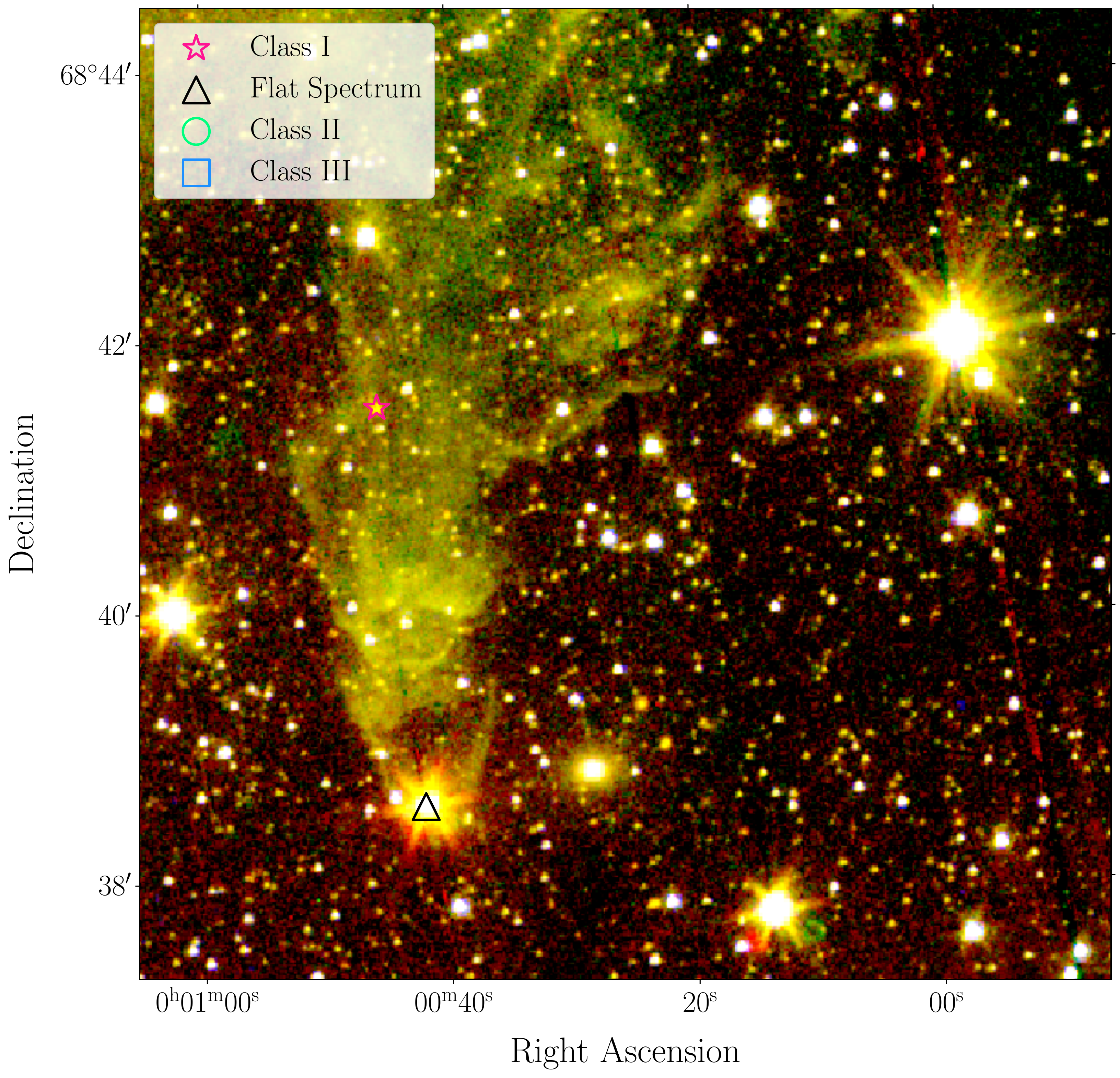}
\includegraphics[width=0.48\linewidth,angle=0]{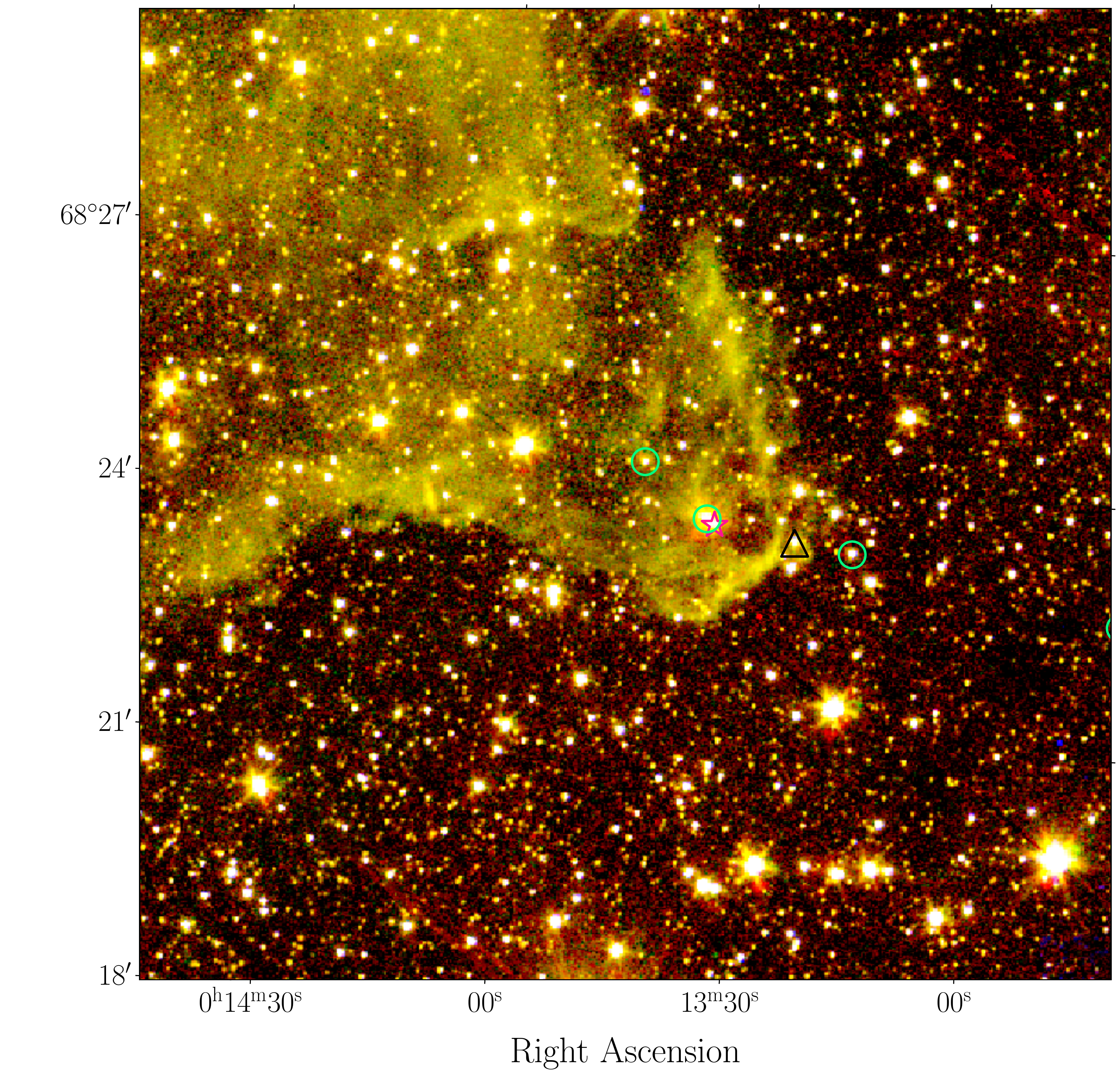}
\caption{Four examples of identified YSOs located in molecular pillars. The YSOs are colored by evolutionary classification and plotted over a color image of 2MASS H-band 1.65 \micron\ (blue), IRAC 3.6~\micron\ (green), and IRAC 4.5~\micron\ (red). The legends in the panels on the left also apply to the panels on the right in the same row. Pillars such as the ones visible in these images form when the ionizing winds from Be~59's massive stars evacuate gas and dust around the cluster, leaving only the dense structures behind. Class I and flat spectrum sources are often seen at the tips of the pillars.\label{fig:pillars}}
\end{center}
\end{figure*}

\begin{figure*}
\begin{center}
\includegraphics[width=0.48\linewidth,angle=0]{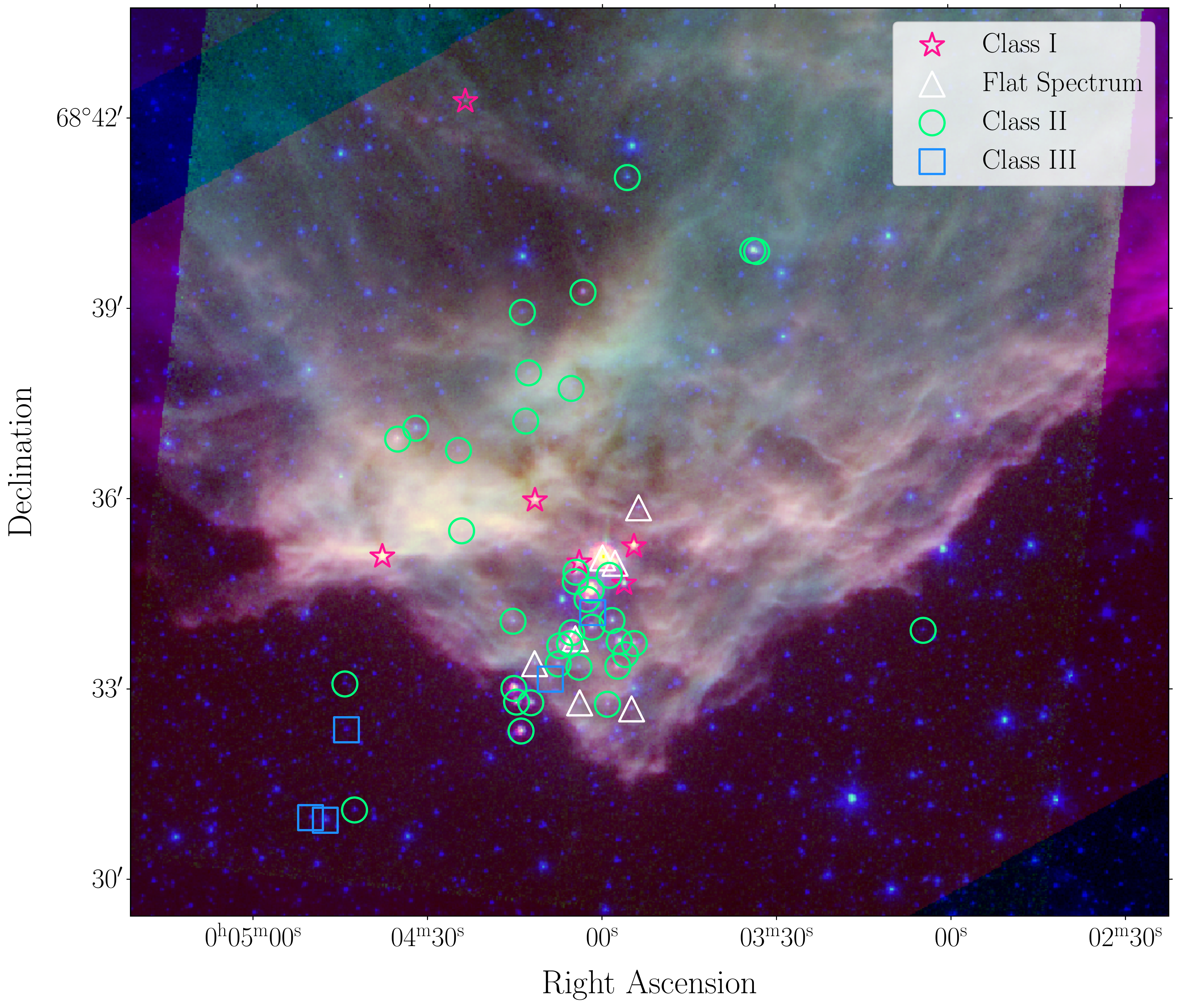}
\includegraphics[width=0.48\linewidth,angle=0]{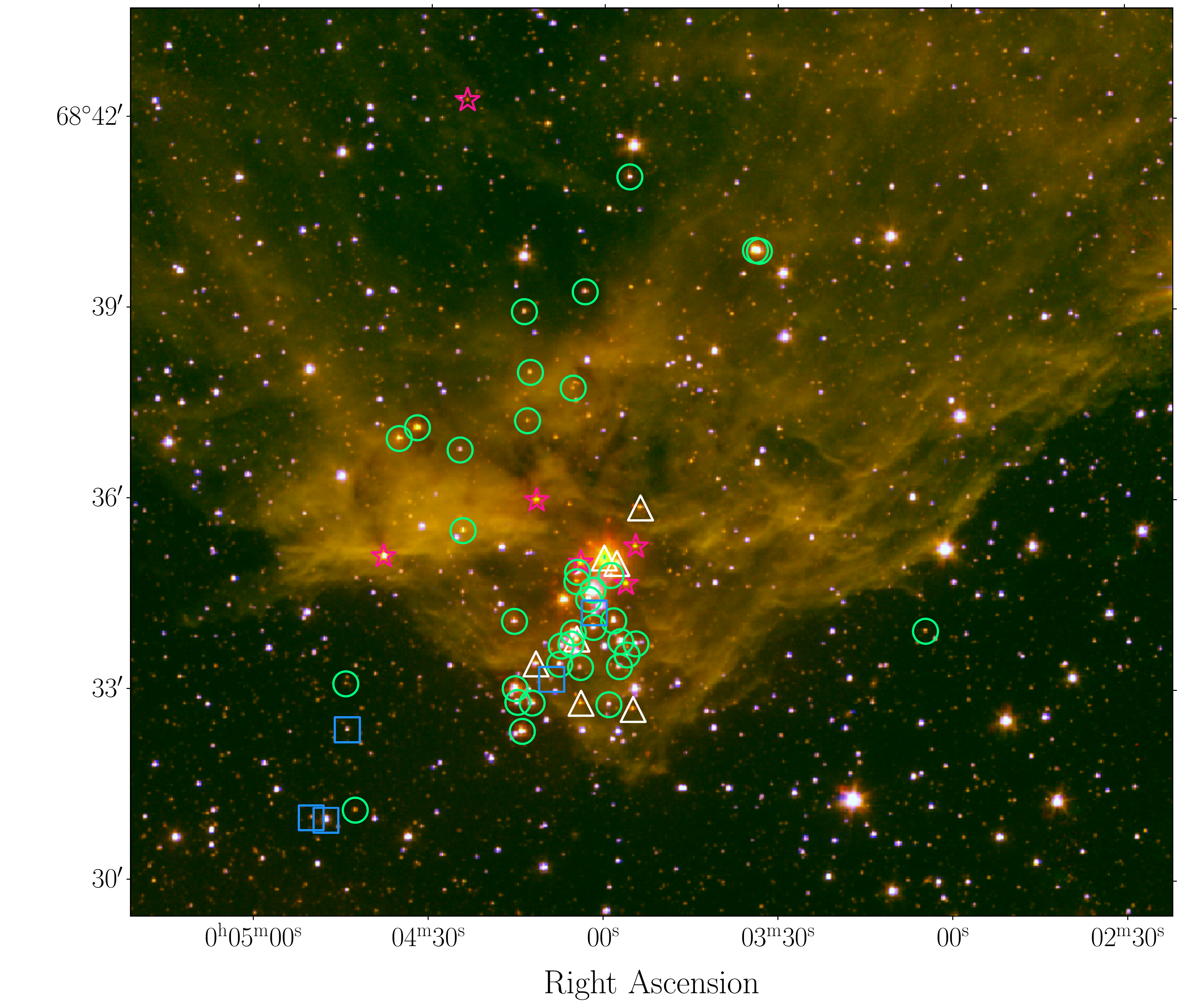}
\caption{Identified YSOs in the northern region around IRAS00013+6817, colored as in Figure~\ref{fig:pillars}. (Left) A color image of IRAC 3.6 \micron\ (blue), IRAC 8.0 \micron\ (green) and MIPS 24 \micron\ (red). (Right) Same as Figure~\ref{fig:pillars} – a color image of 2MASS H-band 1.65 \micron\ (blue), IRAC 3.6~\micron\ (green), and IRAC 4.5~\micron\ (red).  \label{fig:sh171}}
\end{center}
\end{figure*}

\begin{figure*}
\begin{center}
\includegraphics[width=\linewidth,angle=0]{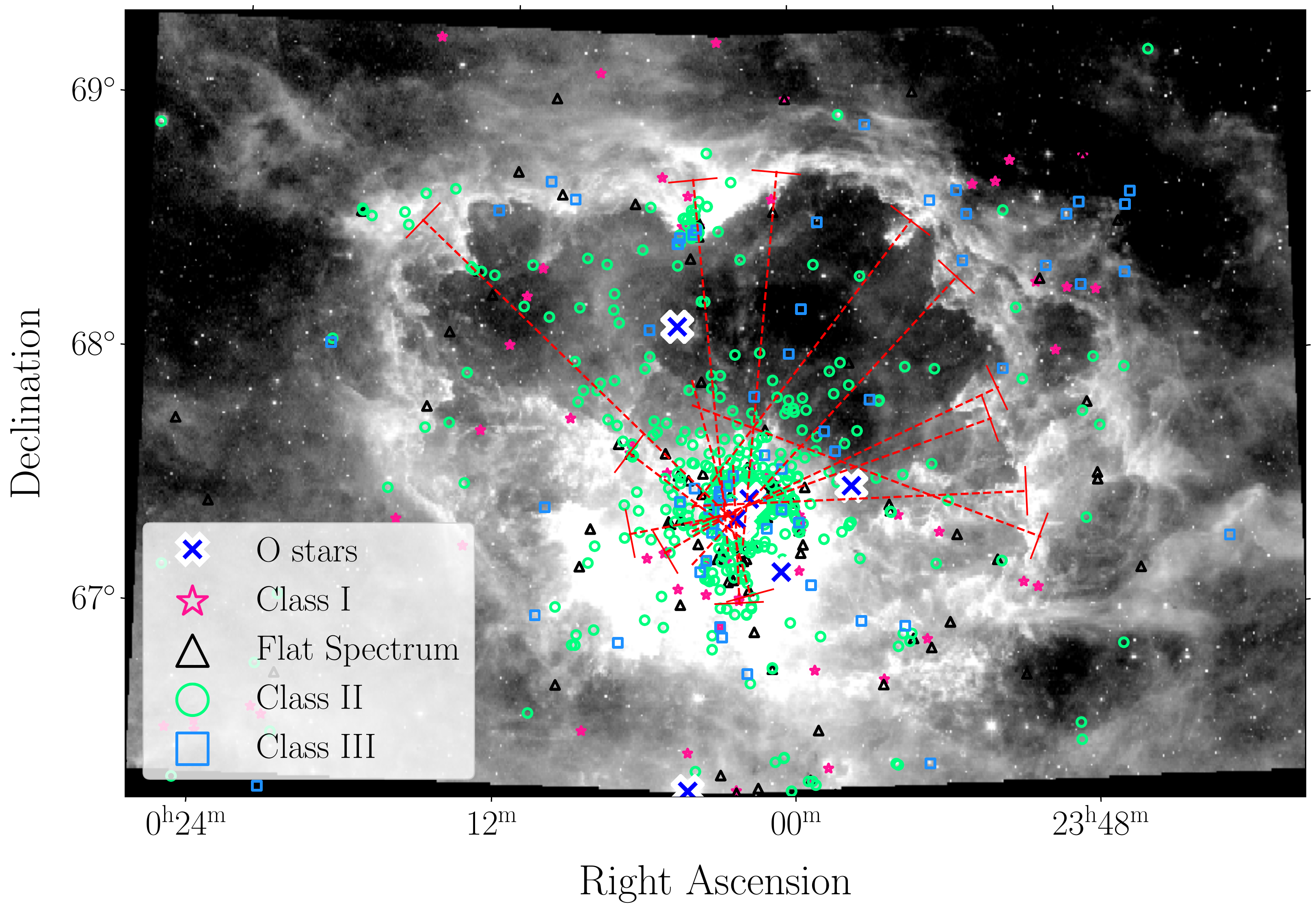}
\caption{The location and direction of some of the pillars in Cep~OB4 plotted over the \textit{WISE} 12 \micron\ image. The dotted red lines begin at the base of the pillars (as indicated by the perpendicular line segment) and follow the pillar orientation to the center of region. The YSOs (including the \textit{WISE}-only YSOs discussed in Section~\ref{subsec:trigger}) are plotted with shapes and colors corresponding to their evolutionary classification as in Figure~\ref{fig:pillars}. The O stars in the region are plotted as blue x's outlined in white. It appears that most of the pillars were exposed by a shockwave originating from Be~59 – likely from one or all of the O stars in that region. \label{fig:pillarsonimage}}
\end{center}
\end{figure*}

\begin{figure*}
\begin{center}
\includegraphics[width=0.32\linewidth,angle=0]{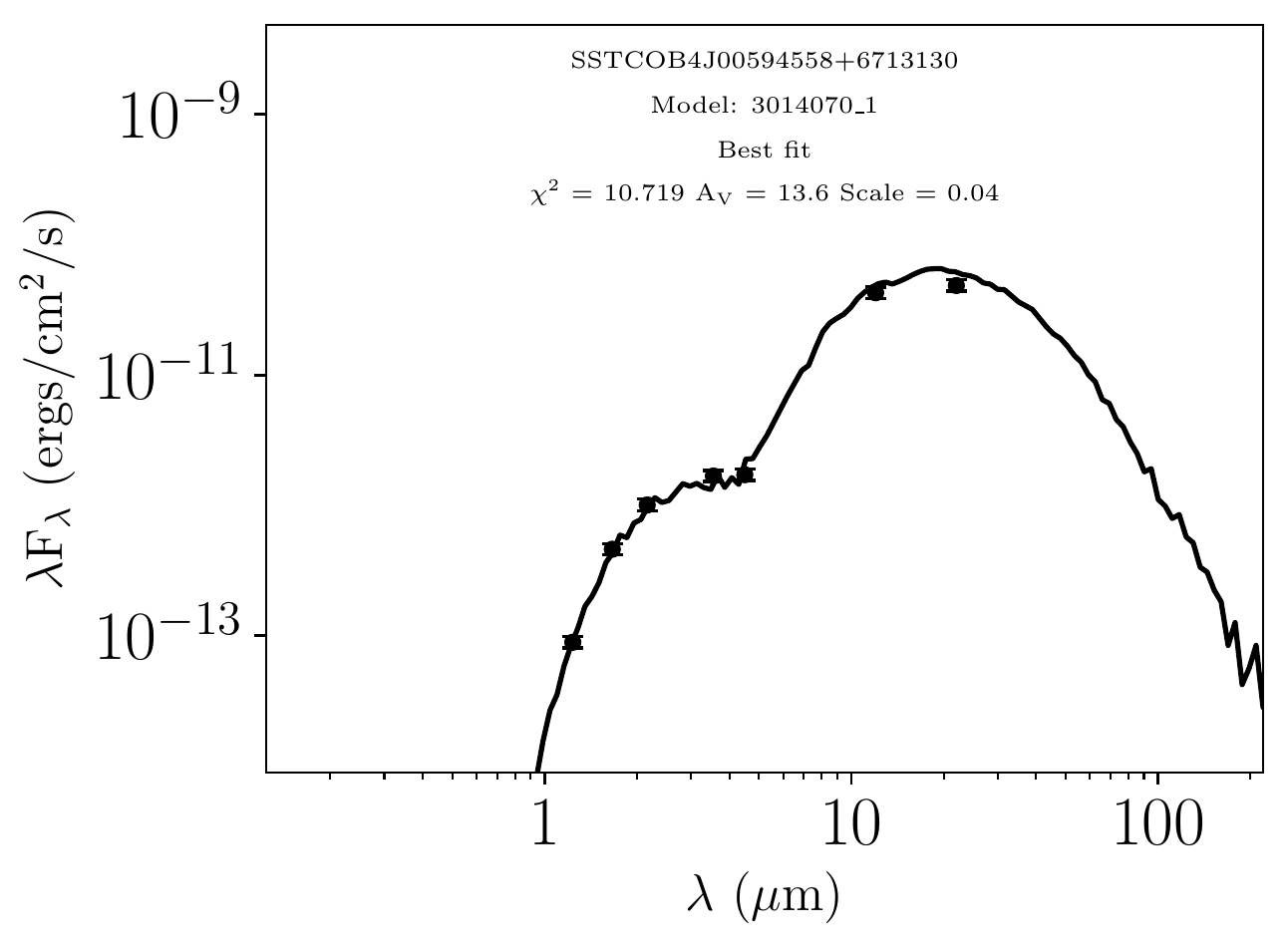}
\includegraphics[width=0.32\linewidth,angle=0]{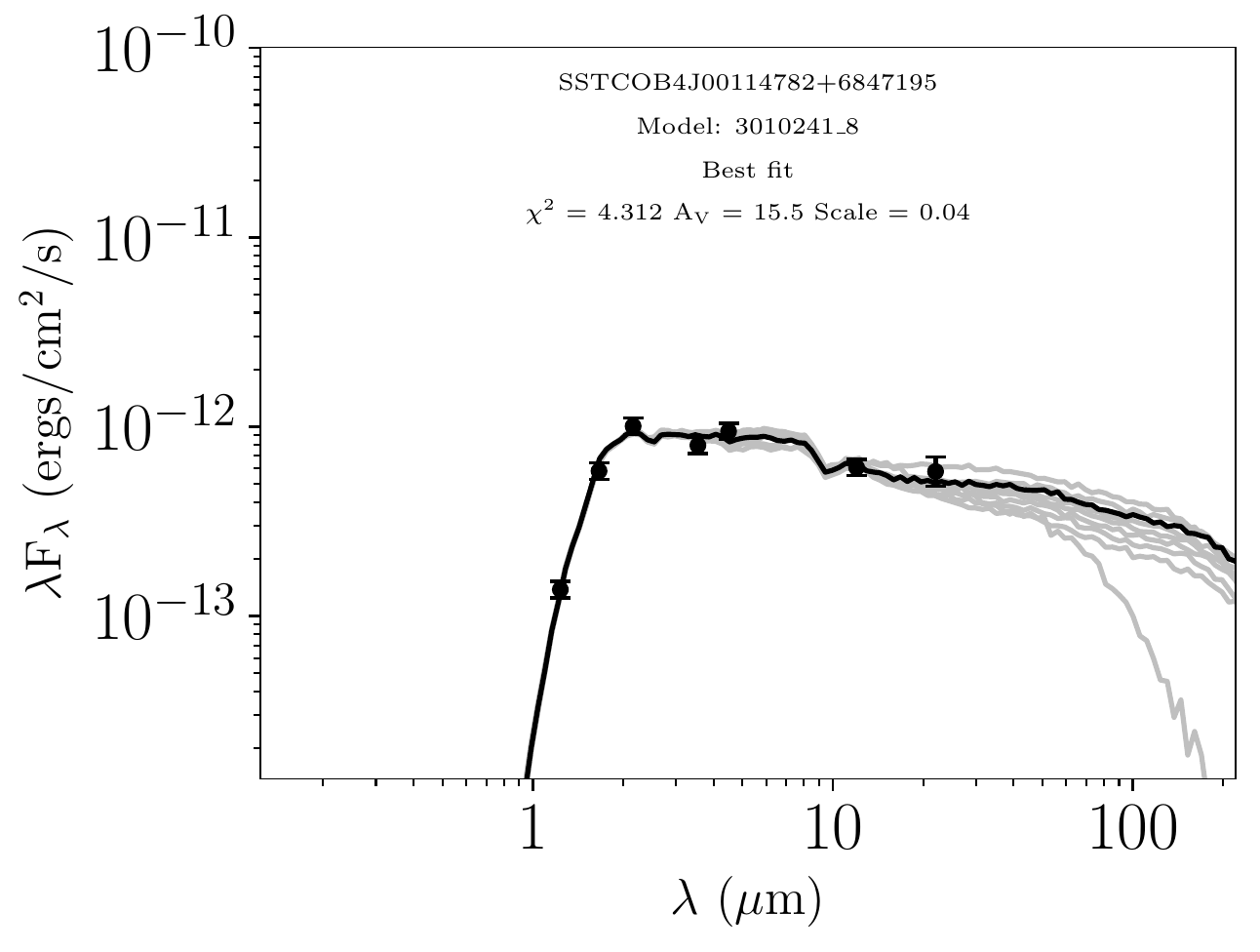}
\includegraphics[width=0.32\linewidth,angle=0]{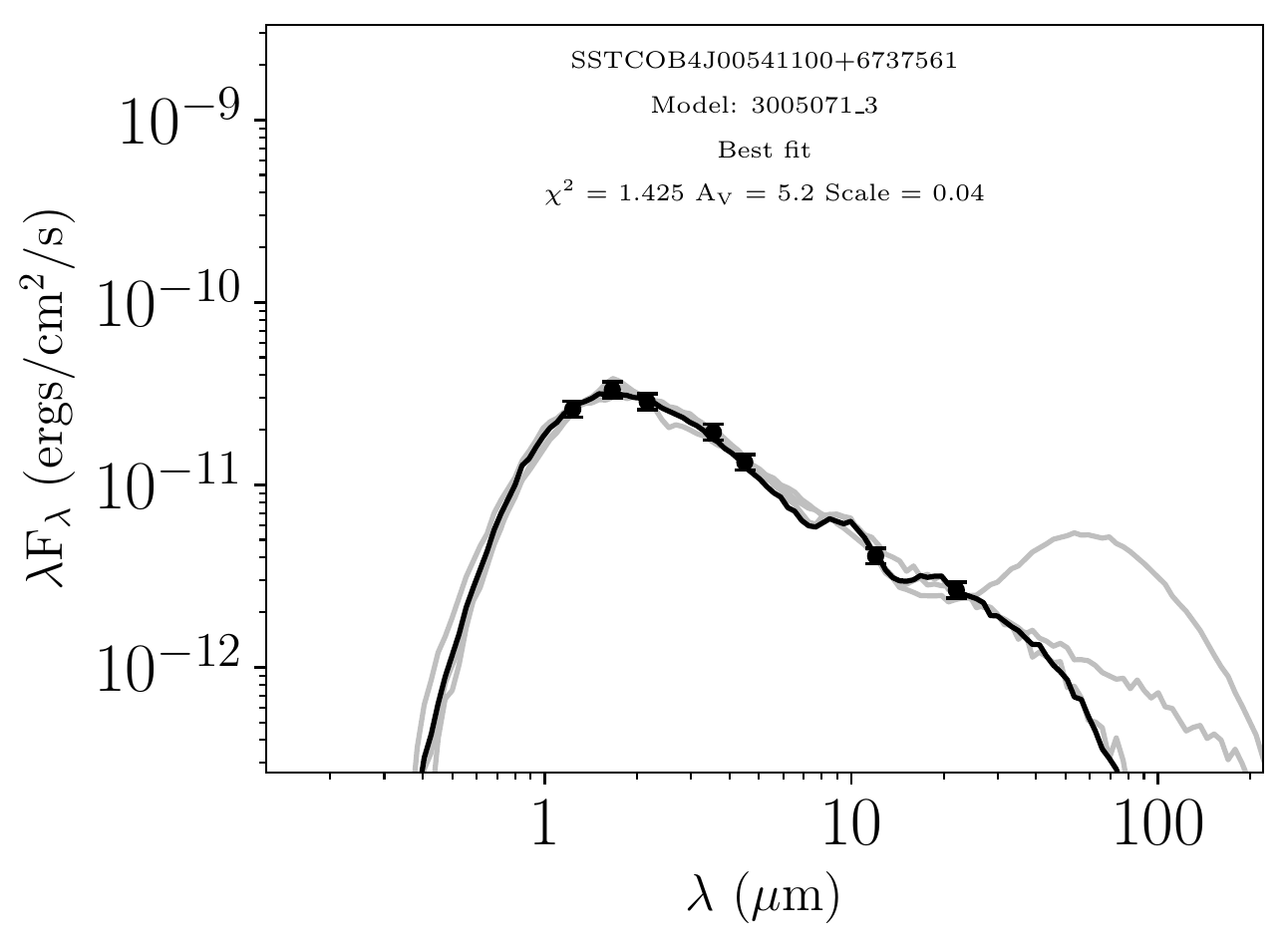}
\caption{Three example fits of YSO SEDs (the complete set of plots of the YSO SED fits are available in the figure set). The leftmost SED is a Class I YSO, the center SED is a flat spectrum YSO, and the rightmost SED is a Class II YSO. The black points represent the photometric data and uncertainty. The black line indicates the best fit model SED and the grey lines show all model fits with $|\chi^2-\chi^2_{best}|<3$.  \label{fig:SEDs}}
\end{center}
\end{figure*}
\begin{figure}
\begin{center}
\includegraphics[width=\linewidth,angle=0]{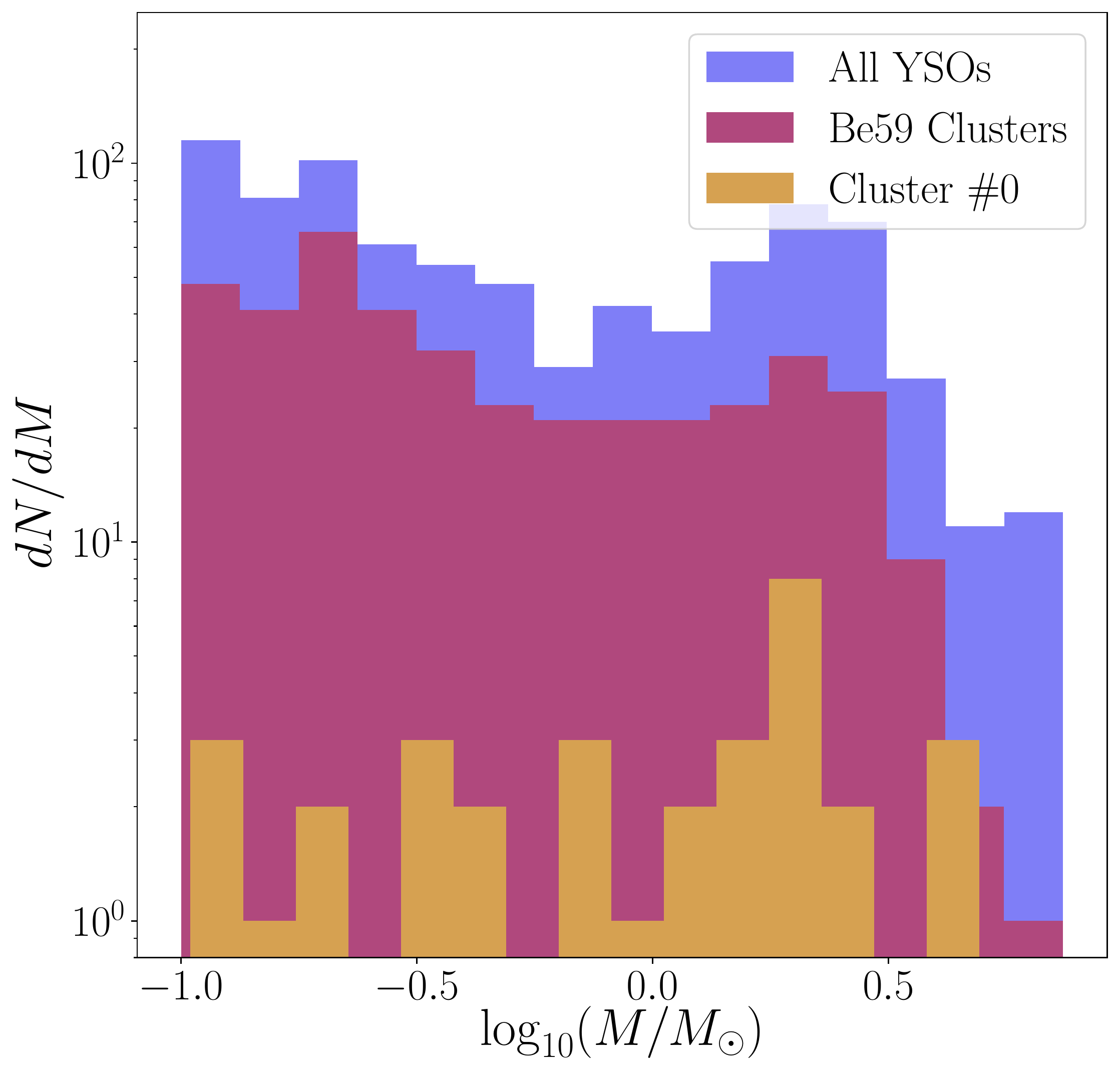}
\caption{The initial mass functions as generated by the SED fitting for all of the YSOs, the YSOs in Cluster \#0, and the YSOs in the clusters near Be~59. We only include SED results with $\chi^2 < 15$. The histograms are overlapping and not normalized. We see the same general trend in each of the samples, a trend that generally agrees with results from previous studies as discussed in Section~\ref{sec:sed}. \label{fig:mass}}
\end{center}
\end{figure}

%\vspace{3em}
\subsubsection{Cluster Properties}\label{subsubsec:properties}
%\vspace{-1em}
Using the parameters determined in Section~\ref{subsubsec:dbscan} we identified 5 clusters of YSOs in Cep~OB4, shown over the IRAC image in Figure~\ref{fig:clusters} and shown individually in Figure~\ref{fig:clusterstogether}. About 63\% of the identified YSOs were assigned to clusters with the remaining 37\% unclustered. The largest cluster size was 445 and the smallest was 6. The median cluster size was 11 and the mean was 103. The locations, sizes, and ratio of Class I and flat spectrum objects to Class II objects are reported in Table~\ref{tab:clusters}. 

The rough area of each cluster was calculated by measuring the convex hull of the cluster members. The convex hull is the polygon formed by using cluster members as vertices so that all cluster members are included in the polygon. The convex hulls of the clusters are plotted in Figure~\ref{fig:clusters}. The circular radius of a cluster is calculated as half the largest separation between two YSOs in the cluster. 

The largest cluster is located in the center of Be\ 59, with two smaller cluster adjacent to it. A mid-sized cluster is located in the northern region in an area of dense nebulosity near IRAS00013+6817. We note that the smallest cluster, Cluster \#4 may be a chance over-density in the field and not a true stellar group. We did not assess the probability of random over-densities rigorously in this study. 

We note that Cluster \#1, the cluster encompassing the majority of Be~59 has the lowest ratio of Class I to Class II objects, indicating that its YSOs are older on average than the other clusters' YSOs. The slightly higher ratio of the nearby clusters – Clusters \#2 and \#3 – suggests that these clusters are younger than the main central cluster and that star formation farther from the central O stars was triggered later. This conclusion is also supported by the higher ratio in the northernmost cluster. We elaborate further on the radial dependence of the Class I to Class II ratio in Section~\ref{subsec:unclusteredysos}. 

\begin{deluxetable*}{crccccc}
\tablecaption{YSO Cluster Properties\label{tab:clusters}}
\tablewidth{0pt}
\tablehead{
\colhead{Cluster} & \colhead{Number} & \colhead{Central RA} & \colhead{Central DEC} &
 \colhead{Circular Radius} & & \colhead{Color in}\\[-0.3cm]
\colhead{Number} & \colhead{of YSOs} & \colhead{(deg)} & \colhead{(deg)} & \colhead{(deg)} & \colhead{Class I/Class II\tablenotemark{a}} & \colhead{Figure~\ref{fig:clusters}}
}
\startdata
0	& 46	& 1.033	& 68.578	& 0.060	& 0.375	& purple \\
1	& 445	& 0.488	& 67.462	& 0.235	& 0.227	& blue \\
2	& 11	& 1.135	& 67.428	& 0.049	& 0.375	& green \\
3	& 6	    & 0.568	& 67.190	& 0.036	& 0.500	& orange\\
4	& 9	    & 1.671	& 67.690	& 0.026	& 0.800	& red
\enddata
\tablenotetext{a}{This ratio is calculated as the number of Class I and flat spectrum objects (as defined in Section~\ref{subsec:evoclass}) over the number of Class II objects.}
\end{deluxetable*}

\begin{deluxetable*}{crcccc}
\tablecaption{YSO Be~59 Subcluster Properties\label{tab:clustersBe59}}
\tablewidth{0pt}
\tablehead{
\colhead{Cluster} & \colhead{Number} & \colhead{Central RA} & \colhead{Central DEC} &
 \colhead{Circular Radius} & \\[-0.3cm]
\colhead{Number} & \colhead{of YSOs} & \colhead{(deg)} & \colhead{(deg)} & \colhead{(deg)} & \colhead{Class I/Class II\tablenotemark{a}}
}
\startdata
0	& 214	& 0.533	& 67.459	& 0.119	& 0.137	\\
1	& 14	& 0.759	& 67.487	& 0.026	& 0.182	\\
2	& 20	& 0.316	& 67.438	& 0.029	& 0.250 \\
3	& 20	& 0.184	& 67.545	& 0.028	& 4.000 \\
4	& 9	& 0.879	& 67.432	& 0.016	& 0.000	\\
\enddata
\tablenotetext{a}{This ratio is calculated as the number of Class I and flat spectrum objects (as defined in Section~\ref{subsec:evoclass}) over the number of Class II objects.}
\end{deluxetable*}

%\vspace{-5em}
\subsubsection{Subclustering in Be~59}\label{subsubsec:subcluster}
To further analyze the YSO distribution in the Be~59 region, we reran the DBSCAN algorithm on the YSOs near Be~59. We selected only YSOs which were within 1.5 times the circular radius of Cluster \#1 as reported in Table~\ref{tab:clusters}. Using the same method described in Section~\ref{subsubsec:dbscan} we obtained DBSCAN parameters of $\epsilon =$ 0\fdg02 \ and MnPts = 9. 
After running the algorithm with these parameters, we identified 5 subclusters in the Be~59 region. We report their properties in Table~\ref{tab:clustersBe59}. We note that Subcluster \#3 has a significantly higher ratio of Class I to Class II objects, indicating that it is younger than the rest of the subclusters and other sources in the central region. We attribute this difference in age to the location of Subcluster \#3. It is situated in an especially dense, nebulous region. The material in the vicinity of Subcluster \#3 was likely not evacuated by the shockwave due to its density, leaving behind a concentration of molecular dust ready to be triggered. We suspect that the O stars which have remained near the center of Be~59 – including BD+66~1673, which is closest in projected distance to the subcluster – are responsible for catalyzing the collapse and the subsequent recent burst of star formation we see in Subcluster \#3.

\subsection{Unclustered YSOs}\label{subsec:unclusteredysos}
As described in Section~\ref{sec:intro}, it is believed that most YSOs form in dense clusters, often impacted by high-energy OB stars. We have identified a large number of YSOs in such clusters in Section~\ref{subsec:clustering}. However, we also expected that many YSOs would not be associated with any groups. There are a number of reasons for the presence of unclustered YSOs. They could have been expelled from their birth cluster through interactions with other cluster members, drifted slowly away from a YSO group as the shockwaves expelled molecular material and changed the gravitational potential of the region, or they might have formed in their current location, never having been associated with a cluster. It is challenging to definitively distinguish between these scenarios without kinematic information on the sources. If we could measure the velocity of a source, we could determine whether it is travelling away from a cluster, perhaps indicating that it was once a member. In the absence of such data, we can only look at the unclustered distribution in aggregate, and make assumptions about the general trajectory of the YSOs.

As described in Section~\ref{subsec:clustering}, we found 63\% of our YSOs associated with clusters. This clustering fraction agrees with previous findings: \citet{Winston2019} found 62\% of YSOs in clusters in the SMOG field, \citet{Winston2020} found 54\% in the region mapped by the GLIMPSE360 program, \citet{Fischer2016} found 53\% in the Cannis Major star-forming region, and \citet{Koenig2008} found between 40\% and 70\% in the W5 \ion{H}{2} region depending on their clustering parameters.

We calculated the density of unclustered YSOs in binned annuli at various distances from the center of Cluster \#1 as reported in Table~\ref{tab:clusters}. The resulting distribution is shown in Figure~\ref{fig:unclustered}. The density of unclustered YSOs is highest in the bins that are closest to the central Be~59 cluster and decreases steeply with increasing projected distance. The density decreases by more than a factor of 2 from the first bin to the second bin - the difference in distance between the bin midpoints being nearly equal to the central cluster's radius. We also calculated the Class~I/Class~II (CI/CII) ratio in the annular bins as a function of distance from the center and show the results in Figure~\ref{fig:unclustered}. The high density of sources at smaller distances and the similarity of the CI/CII ratios to those of the central clusters (a reasonable proxy for relative age) indicate that most of the unclustered YSOs near Be~59 were likely once cluster members themselves. While it is impossible to conclusively determine without velocity information, the age and density of the sources strongly support this conclusion.

The distribution of the CI/CII ratios is especially interesting in the context of triggered star formation. While the distribution of sources is much more complicated than a simple relationship between age and projected distance, the steeply increasing class ratio distribution indicates that (unclustered) YSOs in Cep~OB4 are generally younger farther from the cluster's center. This is evidence in favor of the collect and collapse model of triggered star formation, potentially demonstrating that the expanding shockwave did in fact trigger star formation as it travelled farther from its originating sources in the central cluster.

\subsection{YSOs in Pillars}\label{subsec:pillar}

Another interesting aspect of the spatial distribution of our YSO sample is the presence of numerous YSOs in dense molecular pillars around the border of the \ion{H}{2} region. We show several examples of YSOs in pillars in Figure~\ref{fig:pillars} and YSOs in a dense molecular region in Figure~\ref{fig:sh171}. Such pillars are formed when the shock wave of the expanding \ion{H}{2} region expels unbound dust and gas. What remains around the edges of the bubble are the dense molecular filaments, often sites of star formation. 

The presence of identified YSOs in these locations provides evidence in favor of triggered star formation, specifically for the radiation-driven implosion model described in Section~\ref{sec:intro}. The expanding shockwave likely catalyzed collapse in pre-existing overdensities in Cep~OB4, triggering star formation in exposed molecular structures. In Figure~\ref{fig:pillarsonimage}, we show the approximate origins of the pillars as determined by their direction. It appears as though most of the pillars are pointing towards Be~59, specifically towards the four O stars in the central region: BD+66~1673, LS~I~+67~7, BD+66~1675, and NGC~7822~x. It is not possible for us to distinguish among these O stars due to the imprecision of our estimate of the pillar directions.

We also calculated the projected distance from the base and tip of the pillars to the center of the central cluster as reported in Table~\ref{tab:clusters}. We used an estimate for the shockwave expansion velocity of 15~km~s$^{-1}$ \citep[e.g.][]{Tiwari2021, Luisi2021, Patel1998} to estimate the age of the pillars and the shockwave. Including all of the pillars shown in Figure~\ref{fig:pillarsonimage} resulted in an average expansion time of 1.2 Myr. However, it is important to consider the impact of projection effects. That is, the pillars in the south of the region appear very close to the central cluster even though they may be farther away in actual 3 dimensional space. If we only included the pillars farther to the north – those less affected by the limitations of projection – we calculated an average expansion time of 1.6 Myr. These rough estimates are consistent with Be~59's age of 2 Myr as reported in \citet{Panwar2018}. We also note that there is additional uncertainty in these estimates resulting from the imprecise definition of the ``tip" or ``base" of a pillar. However, the northern pillars had an average projected length of $\approx$3~pc, corresponding to an expansion time of $\approx$0.2~Myr, which would not significantly effect the age estimate of the OB association. The length of the pillars and expansion velocity also provides an upper bound to the age of the YSOs forming in the pillars, assuming they started forming when the expanding shockwave first passed their location. The presence of primarily Class~I and flat spectrum sources in the tips of the pillars (see Figures \ref{fig:pillars} and \ref{fig:sh171}) are also consistent with this estimated age of the pillars.

\section{SED Model Fitting}\label{sec:sed}

In Section~\ref{subsec:evoclass} we used the YSO SEDs to separate the YSOs into evolutionary class. Here, we use the SEDFitter package in Python created by \citet{Robitaille2007} to more rigorously fit the YSO SEDs and obtain mass estimates. The SEDFitter package uses a sample grid of  model SEDs with various ages and masses to fit the YSO distributions. If not specified, the object distance and extinction $A_V$ are treated as free parameters. In our fitting, we allowed $A_V$ to vary between 1 and 50, but set the distance to 1.1 kpc as determined in \citet{Kuhn2019}. 

We crossmatched our YSO catalog with sources from PanSTARRS as described in Section~\ref{subsec:panstarrs}, finding 25 sources with photometry in bands \textit{r} and \textit{i}. We also matched our YSOs with the PMS sources published in \citet{Panwar2018}, which is discussed further in Section~\ref{subsec:previousstudies} and found optical photometry in bands \textit{V} and \textit{I} for 109 of our YSOs. 

We did not include the \textit{WISE} 3.4 and 4.6~\micron\ photometry in the SED fitting as we expect the IRAC 3.6 and 4.5~\micron\ photometry to have more accurately separated sources in crowded regions, and including both would result in unnecessarily inflated $\chi^2$ for some fits. For sources without a \textit{WISE} detection, we set upper limits at 12 and 22~\micron\ of 10 and 8.5 magnitudes respectively, which are the approximate completeness limits of the \textit{WISE} photometry in our sample. For sources with detections at MIPS 24~\micron\, we used only the MIPS 24~\micron\ data and excluded the \textit{WISE} photometry as we considered MIPS to be more reliable because of its higher spatial resolution and better sensitivity. We also used MMIRS photometry when available in place of 2MASS values for the \textit{JHK} bands. Lastly, we required all flux uncertainties to be at least 10\% to take into account possible source variability and the true measurement uncertainties in the 2MASS, IRAC, MIPS, and {\it WISE} datasets. We report the results of the SED fitting in Table~\ref{tab:SED} and show some examples of good SED fits in Figure~\ref{fig:SEDs}. (The complete set of plots of the SED fits is available in the figure set.) 

Because most of our clusters contain a small number of YSOs, we are not able to compare the initial mass function across all clusters. However, we report the IMF for all YSOs, for Cluster \#0, and for the clusters near Be~59 in Figure~\ref{fig:mass}. We find that they are in general agreement with one another and with the IMF of all the YSOs together. While we did not have sufficiently many YSOs to thoroughly compare these IMFs with those from previous studies, we note that the trends we find seem to be in agreement with those found in \citet{Winston2019} and \citet{Winston2020}. While the results reported in Figure~\ref{fig:mass} use only YSOs with $\chi^2<15$ for their SED fit, we report all SED fits in Table~\ref{tab:SED}. 

\begin{figure*}
\begin{center}
\includegraphics[width=0.4\linewidth,angle=0]{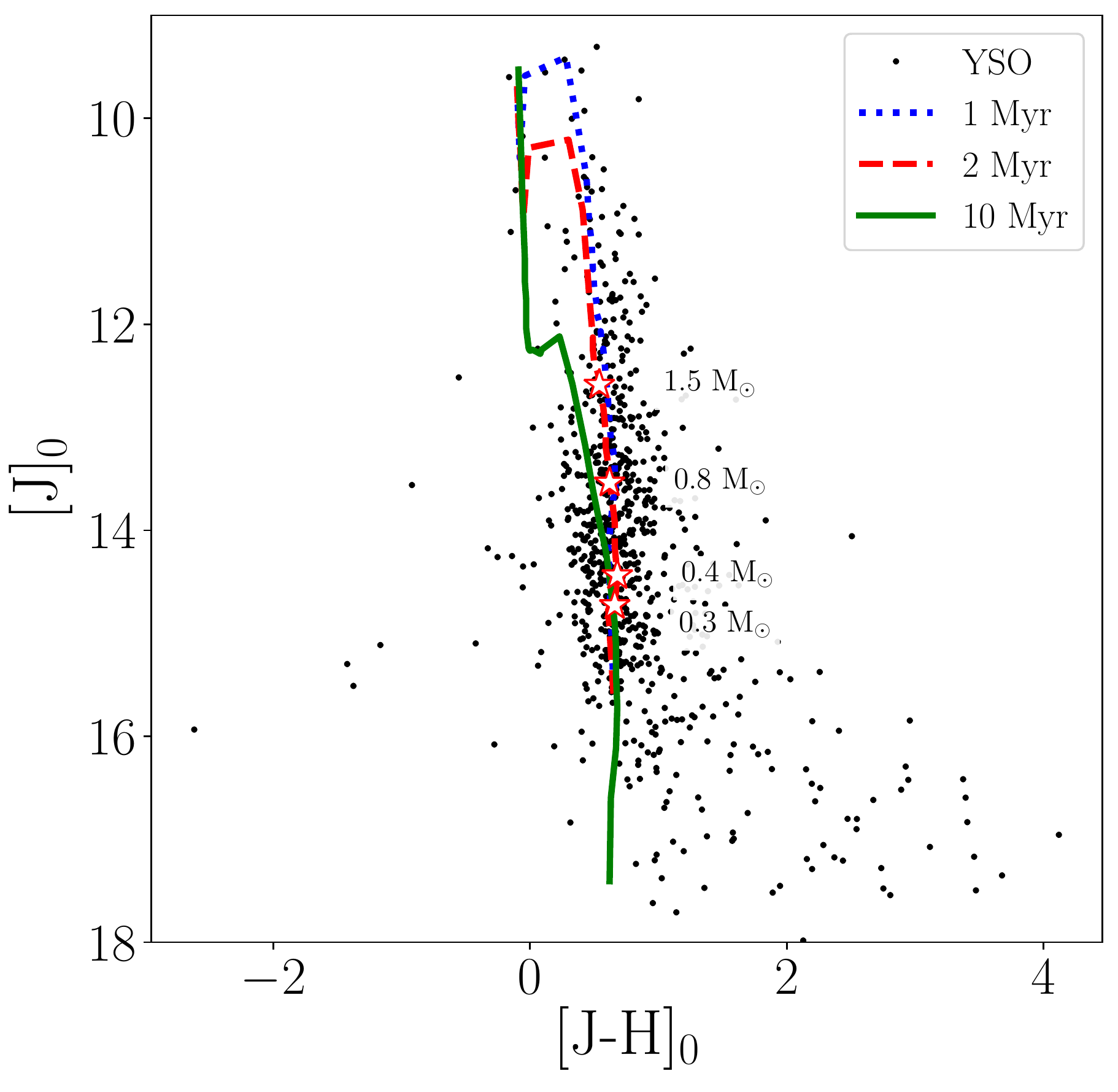}
\includegraphics[width=0.4\linewidth,angle=0]{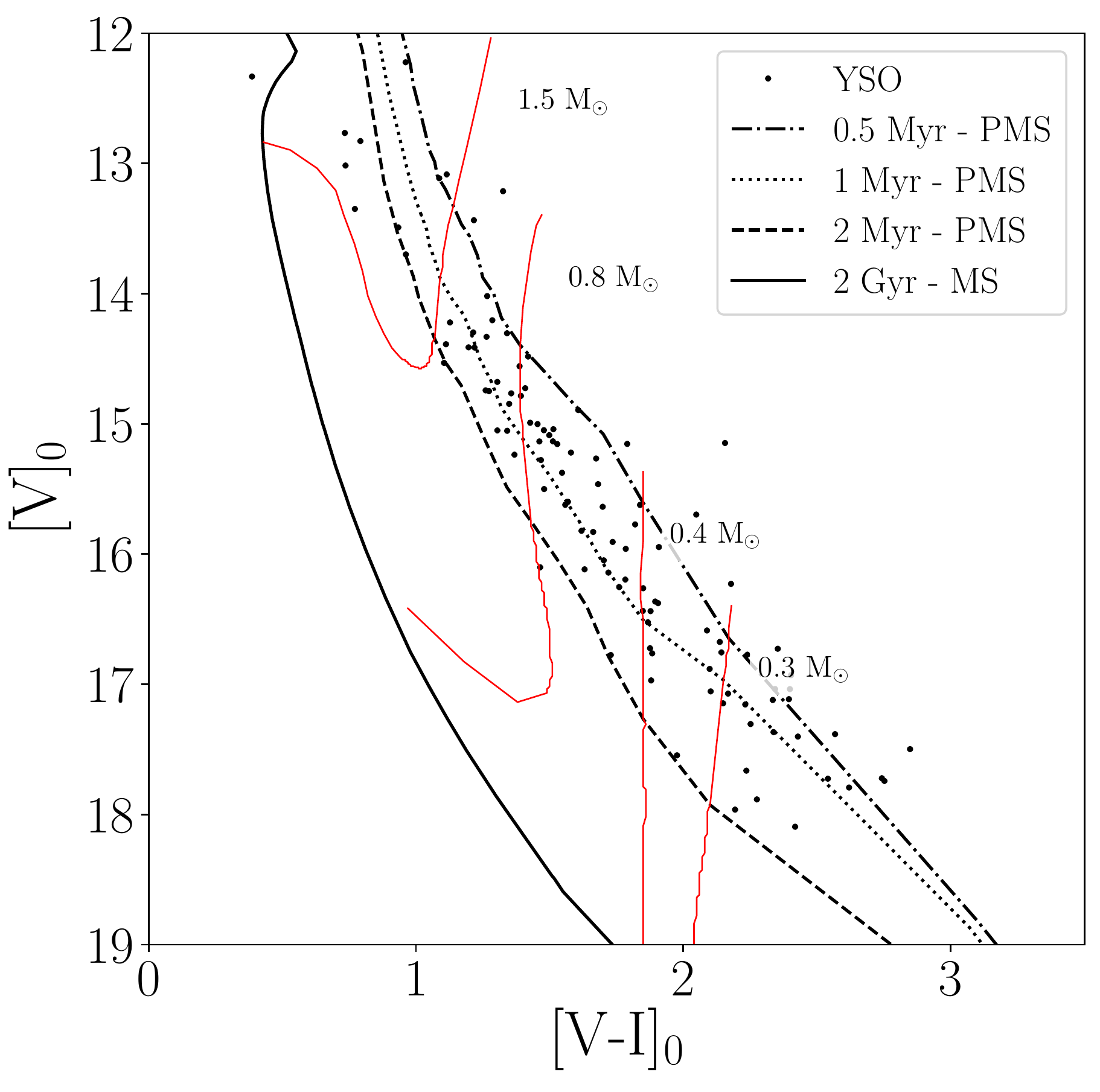}
\caption{Near-infrared (left) and optical (right) color magnitude diagrams of our YSOs. PMS isochrones of various ages from \citet{Siess2000} are plotted in both figures. In the NIR CMD, we show the positions of stars of various masses on the 2 Myr PMS isochrone – shown as white stars outlined in red. In the optical CMD, we show the stars' full evolutionary tracks – shown as solid red curves. A MS isochrone of 2 Gyr from \citet{Girardi2002} is shown in the optical CMD. The YSOs have been dereddened using the NICER extinction maps from \citet{Lombardi2001}. All of the isochrones and evolutionary tracks have been shifted using the assumed distance to Cep~OB4 of 1.1 kpc. \label{fig:cmd}}
\end{center}
\end{figure*}

\begin{figure*}
\begin{center}
\includegraphics[width=0.8\linewidth,angle=0]{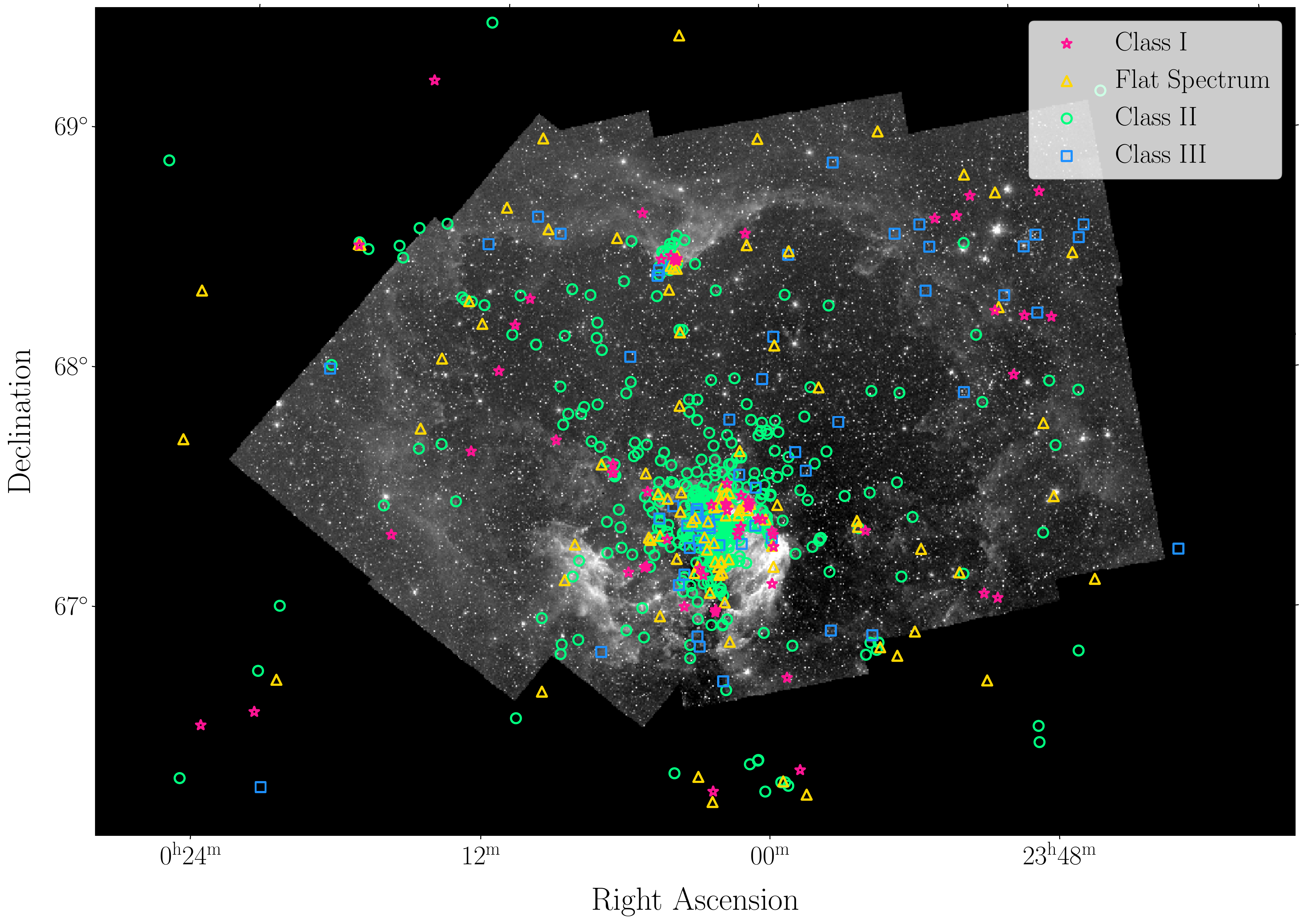}
\caption{The spatial distribution of YSOs in Cep~OB4 plotted over the IRAC 3.6\micron\ image. Here we show the YSOs identified in Section~\ref{sec:YSO} as well as those identified using the larger region queried from the \textit{WISE} catalog as described in Section~\ref{subsec:trigger}. The YSOs are colored by their spectral type. We see numerous young Class I and flat spectrum YSOs towards the outskirts of the cluster and fewer inside the dense border of the \ion{H}{2} region, indicating that there may have been continuing star formation before the creation of the ionized bubble.\label{fig:wiseysos}}
\end{center}
\end{figure*}

\begin{deluxetable*}{cccccccccccc}
\tablecaption{SED Fitting Results\label{tab:SED}}
\tablewidth{0pt}
\tablehead{ 
\colhead{YSO ID} & \colhead{Dist} & 
\colhead{N$_{data}$} & 
\colhead{$\chi^{2}$}   & 
\colhead{M$_C$} & 
\colhead{A$_V$}  &
\colhead{Age} &
\colhead{M$_{disk}$} &  
\colhead{$\dot{M}$}  & 
\colhead{$T_{*}$} & 
\colhead{$L_{*}$}      \\[-0.3cm]
\colhead{} & \colhead{(kpc)}  & 
\colhead{} & 
\colhead{}  & 
\colhead{($M_\odot$)} &  
\colhead{(mag)} &
\colhead{(yr)} &
\colhead{($M_\odot$)} &  
\colhead{($M_\odot$/yr)} &  
\colhead{(K)} &  
\colhead{($L{_\odot}$)} 
}
\startdata
SSTCOB4 J00470610+6745164 	 & 1.1 	 & 6 	 & 12.680 	 & 2.307 	 & 1.000e+00 	 & 1.676e+05 	 & 7.136e-02 	 & 2.178e-04 	 & 4.441e+03 	 & 3.409e+01\\
SSTCOB4 J00473479+6751001 	 & 1.1 	 & 7 	 & 17.790 	 & 0.978 	 & 1.153e+00 	 & 6.726e+05 	 & 5.654e-04 	 & 6.875e-09 	 & 4.250e+03 	 & 2.826e+00\\
SSTCOB4 J00095102+6842133 	 & 1.1 	 & 7 	 & 18.560 	 & 0.118 	 & 5.389e+00 	 & 2.644e+03 	 & 1.101e-04 	 & 2.983e-05 	 & 2.665e+03 	 & 8.307e-01\\
SSTCOB4 J00102105+6845167 	 & 1.1 	 & 5 	 & 1.710 	 & 3.367 	 & 1.841e+00 	 & 8.260e+04 	 & 1.778e-01 	 & 4.992e-04 	 & 4.424e+03 	 & 8.562e+01\\
SSTCOB4 J00114782+6847195 	 & 1.1 	 & 7 	 & 4.312 	 & 0.990 	 & 1.548e+01 	 & 8.258e+06 	 & 1.758e-02 	 & 0.000e+00 	 & 4.246e+03 	 & 5.611e-01\\
SSTCOB4 J00115485+6806090 	 & 1.1 	 & 7 	 & 10.140 	 & 0.221 	 & 4.439e+00 	 & 1.407e+05 	 & 4.765e-04 	 & 7.432e-06 	 & 3.123e+03 	 & 5.596e-01\\
SSTCOB4 J00474717+6748538 	 & 1.1 	 & 5 	 & 5.953 	 & 0.127 	 & 1.238e+00 	 & 6.574e+05 	 & 1.639e-04 	 & 5.343e-08 	 & 2.969e+03 	 & 1.500e-01\\
SSTCOB4 J00484602+6803394 	 & 1.1 	 & 6 	 & 13.580 	 & 2.034 	 & 8.025e+00 	 & 8.854e+05 	 & 9.935e-05 	 & 2.112e-09 	 & 4.705e+03 	 & 6.410e+00\\
SSTCOB4 J00465418+6817350 	 & 1.1 	 & 6 	 & 5.689 	 & 0.122 	 & 1.866e+00 	 & 1.296e+05 	 & 1.092e-03 	 & 9.673e-06 	 & 2.892e+03 	 & 2.937e-01\\
SSTCOB4 J00473180+6818507 	 & 1.1 	 & 6 	 & 36.750 	 & 3.057 	 & 4.580e+00 	 & 1.257e+06 	 & 3.017e-05 	 & 0.000e+00 	 & 5.163e+03 	 & 1.556e+01\\
\enddata
\tablenotetext{}{(This table is available in its entirety in machine-readable form.)}
\end{deluxetable*}
%\vspace{-2em}
\section{Results and Discussion} \label{sec:results}
\subsection{Comparison to Previous Studies}\label{subsec:previousstudies}

\citet{Getman2017} conducted a study of star formation in nearby clouds using X-ray and IR photometry. They employed a classification scheme slightly different than what we described in Section~\ref{subsec:evoclass}. They classified objects as ``disk," ``no disk," or ``possible member of cluster", with ``disk" corresponding to what we describe as Class I, flat spectrum, and Class II, and ``no disk" corresponding to Class III. There is good agreement between our catalogs for the ``disk" sources – we were able to match 92\% of the ``disk" sources to our YSO catalog. The small differences in our catalogs can be attributed to slight differences in photometry and selection criteria. We do not compare the ``no disk" or ``possible member of cluster" sources as they were identified by \citet{Getman2017}  using X-ray photometry. 

We combined our YSO catalog with the additional Getman et al.\ sources and repeated the clustering analysis described in Sections~\ref{subsec:clustering} and \ref{subsubsec:subcluster}. The addition of the extra sources did not significantly affect the YSO distribution or the clustering results. The one notable exception was the detection of an additional cluster in the northern region near our Cluster \#0, where the additional sources added to an existing small group southeast of the cluster (see middle panel of Figure~\ref{fig:clusterstogether}).

\citet{Panwar2018} also studied the Be~59 region, using optical photometry to identify pre-main-sequence stars. We matched the Panwar PMS sources with our YSO catalog using a matching radius of 1\arcsec and found 109 of their 420 sources. There was a significant amount of disagreement between the masses reported in \citet{Panwar2018} and the masses that we derived via SED fitting without including optical photometry. When we included the optical photometry in the V and I bands and reran the SED fitting as described in Section~\ref{sec:sed}, the agreement between masses increased significantly. More work is required to thoroughly identify the source of the discrepancy, but we note that the results of the SED fitting are sensitive to the inclusion of additional photometry. It is not immediately clear whether one set of mass values is more reliable than the other, and it is possible that the disagreement can be attributed to differing methodology and insufficient data. 

We also followed \citet{Panwar2018} and plotted our YSOs along with PMS isochrones and evolutionary tracks from \citet{Siess2000} and MS isochrones from \citet{Girardi2002} in an optical and a NIR color-magnitude diagram in Figure~\ref{fig:cmd}. The NIR CMD is unable to distinguish between the various isochrones, due the isochrones' similarities at dimmer magnitudes and the greater spread in the YSOs' colors at those magnitudes. The optical CMD (while less populated due to the small number of YSOs in our catalog with optical photometry) shows general agreement with an age of 2 Myr for Cep~OB4 as reported in \citet{Panwar2018}. The evolutionary tracks shown illustrate the diversity of masses and evolutionary stages of our identified YSOs as well as predictions for their future evolution. 

\subsection{Triggered Star Formation in Cep~OB4?}\label{subsec:trigger}

If triggered star formation was a significant process in Cep~OB4, ignoring projection effects we would expect to see YSOs generally decreasing in age with increasing distance from the ionizing sources (see Figure~\ref{fig:unclustered}) in the core of the Be~59 cluster with few sources outside of the bubble. We would expect to find significantly more younger Class I and flat spectrum YSOs in the dense border of the \ion{H}{2} region and few to no YSOs beyond the inner edge of the bubble, where the expanding shell has not yet had an effect on the quiescent molecular cloud. Inside the bubble, we would find a large population of Class~II and Class~III objects.
 
The distribution of YSOs in Cep~OB4 in general follows this expectation, as seen in Figure~\ref{fig:wiseysos}.  The Class~II and Class~III objects are heavily clustered around Be~59 and in the region inside of the bubble. There are many Class~I and flat spectrum sources distributed around the edges of the bubble, many at the tips of dense pillars as seen in Figure~\ref{fig:pillars}. The exception to this general pattern is the number of Class~I and flat spectrum sources near Be~59. However, many of these appear to be associated with the dense reflection nebula to the west of the cluster core and the pillars near Be~59, which the stellar winds and expanding ionized shell has not yet dispersed. Because of the variations in the density of the ISM, and the location relative to the ionizing sources, the relative ages and distribution of young YSOs in Cep~OB4 might be more complicated than a simple relation between age and distance from the Be~59 cluster.

To further assess the distribution and evolutionary stages of YSOs beyond the edges of the \ion{H}{2} region, we preliminarily ran the \textit{WISE} YSO selections described in Section~\ref{subsec:WISEysosel} and Appendix~\ref{appendix:WISE} on the entire \textit{WISE} catalog queried as described in Section~\ref{subsec:WISE}. We identified 169 additional YSOs not found in the selection described in Section~\ref{sec:YSO}. All of these YSOs were located around the border or beyond the area covered by the IRAC images. The distribution of the YSO from the catalog compiled in Section~\ref{sec:YSO} and the 169 additional \textit{WISE} YSOs is shown in Figure~\ref{fig:wiseysos}.

From a cursory visual analysis of the distribution, we see that there are numerous young Class I and flat spectrum YSOs just outside of the border and beyond the \ion{H}{2} region. The presence of young YSOs – and therefore ongoing or recent star formation – outside of the \ion{H}{2} region and beyond the influence of the ionizing shock-wave indicates that there was likely continuing star formation before the creation of the ionized bubble. These objects could also be foreground or background star forming regions not associated with Be~59 and the Cep~OB4 cloud. We would have to know the 3-D space location of the YSOs within and outside of the bubble to have a full picture of the star formation sequence in the region. However, the distances to the YSOs are difficult to obtain. They are in general optically faint and undetected by {\it Gaia} or other parallax observations, and their distance also adds to the difficulty of these measurements.

\section{Conclusions}\label{sec:conclusion}

We have conducted a study of YSOs in the Cep~OB4 OB association, using new \textit{Spitzer} IRAC and MMIRS images, publicly available MIPS images, and photometry from 2MASS and \textit{WISE}. 

\begin{itemize}
    \item We extracted 752,615 total unique sources from the IRAC 3.6 and 4.5~\micron\ mosaics. From the limited area mapped by IRAC 5.8 \micron\, 8.0 \micron, and MIPS 24 \micron, we extracted 2,151, 1,641, and 485 sources respectively with corresponding photometry in IRAC 3.6 or 4.5~\micron. We also extracted 113,424 H-band, 114,118 J-band, and 18,369 K-band sources from the MMIRS mosaics that had photometry in IRAC 3.6 or 4.5~\micron.
    \item We cross matched our catalog with photometry from \textit{WISE} and 2MASS, finding 140,594 sources with 2MASS photometry and 126,835 sources with \textit{WISE} photometry.
    \item We identified 821 YSOs with excess IR emission using a variety of color-color cuts.
    \item We classified the YSOs based on their IR SED slope $\alpha_{IR}$ and found 67 Class I,  103 flat spectrum sources, 569 Class II, and 82 Class III.
    \item We identified 5 YSO clusters, 3 located in the Be~59 region and 1 in the region around IRAS00013+6817. The median cluster size was 11 YSOs and the mean was 103.
    \item We identified many YSOs located in dense molecular pillars around the edges of the \ion{H}{2} region.
    \item We fit the YSO SEDs with YSO models to estimate their masses and other physical parameters. We constructed the IMF for Cep~OB4 and found it to be in general agreement with previous studies.

\end{itemize}

The distribution of Class~I and flat spectrum YSOs is in general consistent with the triggered star formation scenario, but further observations are necessary to confirm that this model describes the sequence of star formation in this region. We plan to propose for Submillimeter Array \citep[][]{Ho2004} observations to find younger Class~0 and deeply embedded Class~I YSOs that are not detected in the IRAC images to locate active sites of star formation in the region.
Kinematic data that will allow us to determine the 3D positions and velocities of the stars in the region will help us determine cluster membership and dynamics. We will also perform near-IR spectroscopy on a sample of the YSOs we have detected in order to better determine their masses and ages, which will provide a clearer picture of the star formation activity in Cep~OB4.

%
%A handy "cheat sheet" that provides the necessary LaTeX to produce 17 
%different types of tables is available at %\url{http://journals.aas.org/authors/aastex/aasguide.html#table_cheat_sheet}.




% \clearpage
\vspace{5mm}
%\begin{acknowledgments}    %commented out otherwise it adds line numbers 
This work is based on observations made with the {\it Spitzer} Space Telescope, which was operated by the Jet Propulsion Laboratory, California Institute of Technology 
under NASA contract 1407. 
Support for the IRAC instrument was provided by NASA through contract 960541 issued by JPL. 

This publication makes use of data products from the Two Micron All Sky Survey, which is a joint project of the University of Massachusetts and the Infrared Processing and 
Analysis Center/California Institute of Technology, funded by the National Aeronautics and Space Administration and the National Science Foundation. 

This publication makes use of data products from the Wide-field Infrared Survey Explorer, which is a joint project of the University of California, Los Angeles, and the Jet Propulsion Laboratory/California Institute of Technology, funded by the National Aeronautics and Space Administration.
This research made use of Montage. It is funded by the National Science Foundation under Grant Number ACI-1440620, and was previously funded by the National Aeronautics 
and Space Administration's Earth Science Technology Office, Computation Technologies Project, under Cooperative Agreement Number NCC5-626 between NASA and the 
California Institute of Technology.
This research has made use of NASA’s Astrophysics Data System.

The Pan-STARRS1 Surveys (PS1) and the PS1 public science archive have been made possible through contributions by the Institute for Astronomy, the University of Hawaii, the Pan-STARRS Project Office, the Max-Planck Society and its participating institutes, the Max Planck Institute for Astronomy, Heidelberg and the Max Planck Institute for Extraterrestrial Physics, Garching, The Johns Hopkins University, Durham University, the University of Edinburgh, the Queen's University Belfast, the Center for Astrophysics $|$ Harvard \& Smithsonian, the Las Cumbres Observatory Global Telescope Network Incorporated, the National Central University of Taiwan, the Space Telescope Science Institute, NASA Grant No. NNX08AR22G issued through the Planetary Science Division of the NASA Science Mission Directorate, the NSF Grant No. AST-1238877, the University of Maryland, Eotvos Lorand University (ELTE), the Los Alamos National Laboratory, and the Gordon and Betty Moore Foundation.

The SAO REU program is funded in part by the National Science Foundation REU and Department of Defense ASSURE programs under NSF Grant no.\ AST-1852268, and by the Smithsonian Institution. 
%\end{acknowledgments}

\facilities{Spitzer (IRAC), WISE}
\software{Astropy \citep{Astropy2013,Astropy2018}, SAOimageDS9 \citep{Joye2003}, Montage \citep{Berriman2008}, SEDFitter \citep{Robitaille2007}}

\appendix

\section{Source Extraction and Photometry}
\subsection{IRAC and MIPS Photometry}\label{appendix:ExtractPhotoIRAC}
We performed photometry on the IRAC band 1, 2, 3, and 4 mosaics as well as the MIPS 24~\micron\ image using {\tt Source Extractor} \citep[SExtractor;][]{Bertin1996}.  We repeated the source extraction for the short frame mosaics as well in order to obtain accurate magnitudes for the brighter sources. We adjusted the SExtractor parameters to ensure that most obvious sources were identified while minimizing the number of false detections. The background parameters BACK\_SIZE and BACK\_FILTERSIZE were the most influential parameters in this process. Setting either to a value much larger than the average source size would neglect variations in background across the image (specifically in regions with significant nebulosity). Setting either parameter too small would cause the source flux to be included in the background, potentially preventing their detection. The SExtractor parameters are reported in Table~\ref{tab:sextractor} for each of the mosaics. 

The zero-point magnitudes were determined for each band using observations of a set of standard stars. For bands 1 and 2 we used standard stars KF06T1, KF06T2, KF08T3, KF09T1, NPM1p60.0581, NPM1p67.0536, and NPM1p68.0422; for bands 3 and 4 we used standard stars NPM1p64.0581, HD165459, NPM1p66.0578, NPM1p67.0536, and NPM1p68.0422. The images of the IRAC calibration stars were obtained by the IRAC calibration program in 2019 January (near in time to the Program 14005 AORs). We produced mosaics of the standard stars using the same methods as used for the Cep~OB4 mosaics. For the MIPS calibration, we used calibration stars HD 159330 and HD 173398 whose magnitudes are reported in \citet{Engelbracht2007}. SExtractor was run on each standard star using the parameters determined for the Cep~OB4 long frame mosaics as described above. The zero-point magnitudes for each band were calculated by minimizing the mean of the difference between the measured magnitudes of the standard stars and their reported calibration magnitudes in \citet{Reach2005} or \citet{Engelbracht2007}. After the zero-point magnitudes were determined, we ran SExtractor again on the Cep~OB4 mosaics to produce the final calibrated IRAC catalogs and the MIPS catalog. All of the reported IRAC and MIPS magnitudes are calibrated based on \citet{Reach2005} and \citet{Engelbracht2007} respectively, which both base their absolute calibration on Vega. 

\subsection{MMIRS Photometry}\label{appendix:ExtractPhotoMMIRS}
The MMIRS mosaics at each map pointing were extracted separately and the catalogs with calibrated photometry were merged at the end to eliminate duplicate sources where the mosaics overlapped along the edges of each FOV. There was some residual nonlinearity in the extracted photometry, and so the photometry could not be calibrated with a single zero point magnitude. For each image, we cross matched the extracted MMIRS sources to the 2MASS catalog and fit a line to the magnitude difference versus MMIRS magnitude for values with 2MASS magnitudes between 13 and 15 mag. We took the average slope of all images in each band in each set of images (SAO-8-19c and SAO-12-21a) and refit the models using the fixed average slopes but allowing the vertical offset to vary between images. The values of the average slope values are reported with the SExtractor parameters in Table~\ref{tab:sextractormmirs}. The mosaics were calibrated separately to correct for airmass and to remove the effects of variable sky transmission.

\section{2MASS+IRAC YSO Selection}\label{appendix:2MASSIRAC}

We identified and removed the background sources described in Section~\ref{subsec:2MASScontaminants} using our IRAC 3.6 and 4.5~\micron\ photometry. We followed the selections described in \citet{Winston2020}, which we reformat here for clarity. Objects were considered background sources if they met one of the conditions specified in Equation \ref{eq:2MpIRAC1}:
\begin{equation}\label{eq:2MpIRAC1}
\begin{split}
    [3.6] < 6.0 \mbox{  or  } \\
    [4.5] < 5.5 \mbox{  or  } \\
    ([3.6]>16.0  \mbox{  and  } [3.6-4.5] \geq 1.5) \mbox{  or  }\\
    ([3.6]>14.0  \mbox{  and  } [3.6-4.5]\leq 0.5) \mbox{  or  } \\  ([3.6]>2([3.6-4.5]+0.5)+14 \mbox{  and  } [3.6-4.5]>0.5 \mbox{  and  } [3.6-4.5]<1.5)
\end{split}
\end{equation}

Our catalog contained 752,615 total sources 711,152 (94\%) of which were identified as background sources. This left a remaining 41,463 YSO candidates, 41,059 of which had corresponding 2MASS photometry as determined in Section~\ref{subsec:crossmatch}. The results of this background source removal are shown in Figure~\ref{fig:contaminants}.

We then calculated the extinction coefficients $A_{K_s}$ following the method of \citet{Flaherty2007}. We only considered the sources with valid values of $A_{K_s}$, which eliminated 29 sources and reduced the number of YSO candidates to 41,030. We dereddened the photometry for $[K_s-3.6]$ and $[3.6-4.5]$ as follows:

\begin{equation}\label{eq:2MpIRAC2}
    \begin{split}
        [3.6 - 4.5]_0 = ([3.6]- A_{3.6}\cdot A_{K_s}) -([4.5]- A_{4.5} \cdot A_{K_s})\\
        [{K_s} - 3.5]_0 =   ([{K_s}]-  A_{K_s}) - ([3.6]- A_{3.6} \cdot A_{K_s})
    \end{split}
\end{equation}
Where the color excess coefficients are given as $A_j =2.5$, $A_H=1.55$, $A_{3.6}=0.63$, and $A_{4.5}=0.53$.

We then used three different combinations of 2MASS and IRAC photometry to identify the YSOs from the remaining candidates. We  based our selection criteria on that of \citet{Winston2020}, but adjusted the criteria in (\ref{eq:2MpIRAC4}) to better fit our data. We adjusted the selection to be less conservative to account for the smaller amount of scatter in our color-color distribution. In each selection, only sources with valid uncertainty values were used. In the following expressions, we calculate the color uncertainties as $\sigma_{AB}^2 = \sigma_A^2 + \sigma_B^2$.

\begin{equation}\label{eq:2MpIRAC3}
\begin{split}
\frac{A_H -A_{4.5}}{A_J -A_H}([J-H]-0.6+\sigma_{JH})+1.0+ \sigma_{H4.5} <[H-4.5]  \mbox{   and   }\\
[J - H] > 0
\end{split}
\end{equation}

\begin{equation}\label{eq:2MpIRAC4}
\begin{split}
\frac{A_{K_s} -A_{4.5}}{A_H -A_{K_s}}([H-{K_s}]+\sigma_{H{K_s}})+0.4+\sigma_{{K_s}4.5} < [{K_s}-4.5]  \mbox{   and   }\\
[H - {K_s}] > 0 \mbox{   and   }\\
[{K_s} - 4.5] > 0.2 + \sigma_{{K_s}4.5}
\end{split}
\end{equation}

\begin{equation}\label{eq:2MpIRAC5}
\begin{split}
[3.6 - 4.5]_0 - \sigma_{3.6,4.5} > 0 \mbox{  and  } \\ 
[{K_s} - 3.6]_0 - \sigma_{{K_s}3.6} >0.2\cdot[3.6-4.5]+0.3  \mbox{  and  } \\
[{K_s} - 3.6]_0 - \sigma_{{K_s}3.6} > -1.0([3.6 - 4.5]_0 - \sigma_{12}) + 0.8
\end{split}
\end{equation}

After implementing these selections, we found 388 YSOs using selection (\ref{eq:2MpIRAC3}), 514 using selection (\ref{eq:2MpIRAC4}), and 656 using selection (\ref{eq:2MpIRAC5}) for a total of 719 YSOs identified using the combined 2MASS and IRAC photometry (see Figure~\ref{fig:2MASSyso}). 

\section{WISE YSO Selection}\label{appendix:WISE}

Again following the methods of \citet{Winston2020} based on \citet{Fischer2016} and \citet{Koenig2014}, we removed background sources in the subset of our data with \textit{WISE} photometry and identified YSOs from the remaining candidates.

We used the following criteria for identifying background AGN and SFG:
\setcounter{equation}{0}
\begin{equation}
\begin{split}
  \mbox{SFG} =[W_2 - W_3] > 2.3 \mbox{  and  }\\
    [W_1 - W_2] < 1.0 \mbox{  and  }\\
    [W_1 - W_2] < 0.46([W_2 - W_3] - 0.78)  \mbox{  and  }\\
    [W_1] > 14
\end{split}
\end{equation} 
\begin{equation}
\begin{split}
  \mbox{AGN} =[W_1] > 1.8([W_1 - W_3] + 4.1) \mbox{  and  }\\
    [W_1] > 14 < 1.0 \mbox{  or  }\\
    [W_1] > [W_1 - W_3] + 10.0
\end{split}
\end{equation}    

Of the 126,835 sources with \textit{WISE} photometry, 116,796 were identified as AGN and 68,634 as SFG. This left 6,526 remaining \textit{WISE} YSO candidates.

To select the YSOs from the candidates, we used the criteria described in \citet{Fischer2016}. Class 1 \textit{WISE} YSOs satisfy:

Class I \textit{WISE} YSOs satisfy all four of the following constraints:
\begin{equation}
    \begin{split}
        W_2 - W_3 > 2.0 \mbox{  and  }\\
        W_2 - W_3 < 4.5\mbox{  and  }\\
        W_1 - W_2 >0.46 \cdot(W_2-W_3)-0.9\mbox{  and  }\\
        W_1 - W_2 > -0.42 \cdot (W_2 - W_3) + 2.2
    \end{split}
\end{equation}

And Class II \textit{WISE} YSO satisfy: 
\begin{equation}
    \begin{split}
        W_1 - W_2 > 0.25\mbox{  and  }\\
        W_1 - W_2 < 0.71 \cdot (W_2 - W_3) - 0.07\mbox{  and  }\\
        W_1 - W_2 > -1.5 \cdot (W_2 - W_3) + 2.1\mbox{  and  }\\
        W_1 - W_2>0.46\cdot(W_2-W_3)-0.9\mbox{  and  }\\
        W_1 - W_2 < -0.42 \cdot (W_2 - W_3) + 2.2
    \end{split}
\end{equation}

Using these selections, we identified 235 total \textit{WISE} YSOs, 52 Class I and 183 Class II (see Figure~\ref{fig:WISE}).

\section{WISE+IRAC Source Selection}\label{appendix:WISEIRAC}

The \textit{WISE}+IRAC source selection was preformed in exactly the same manner as the selection described in Appendix~\ref{appendix:WISEIRAC}, but using IRAC 3.6 and 4.5~\micron\ photometry instead of \textit{WISE} bands 1 and 2. 
Of the 126,835 sources with \textit{WISE} photometry, 111,879 were identified as AGN and 59,491 as SFG. We also removed 5,625 additional sources that were identified as background sources using 2MASS+IRAC photometry in Appendix~\ref{appendix:2MASSIRAC}. This left 5,217 remaining \textit{WISE}+IRAC YSO candidates.

We identified 196 total \textit{WISE}+IRAC YSOs, 22 Class I and 174 Class II.

\section{MMIRS+IRAC Source Selection}\label{appendix:MMIRSIRAC}
The MMIRS+IRAC YSO selections was performed using the same background source removal and selection criteria as in the 2MASS+IRAC selection described in Appendix~\ref{appendix:2MASSIRAC}. We only included sources with MMIRS photometry below the saturation limit: 12.7 magnitudes for H-band, 12.5 magnitudes for J- band, and 11.1 magnitudes for K-band. We also excluded sources that had corresponding 2MASS photometry above these limits as such sources were saturated and thus had unreliable photometry in the MMIRS images.

After the background source removal, we were left with 10,231 sources with good MMIRS photometry in at least one band, one of which was removed because of a bad extinction coefficient. Notably, due to limited coverage, there were only 1,408 field sources with good K-band MMIRS photometry. 

After applying the selections from Appendix~\ref{appendix:2MASSIRAC}, we found 227 YSOs using selection (\ref{eq:2MpIRAC3}), 10 using selection (\ref{eq:2MpIRAC4}), and 3 using selection (\ref{eq:2MpIRAC5}) for a total of 232 YSOs. This distribution of selection method is due to the very limited K-band MMIRS coverage. 

\section{IRAC Bands 3 and 4 Selection}\label{appendix:IRAC34}
\setcounter{equation}{0}
In the regions covered by our IRAC band 3 and/or band 4 mosaics, we removed background sources and selected YSOs following the methods of \citet{Winston2019}. We selected 1,166 AGN candidates using the following criteria:

\begin{equation}
    \begin{split}
        AGN_1 = [4.5-8.0]>1.2\mbox{  and  }\\
        [4.5] > 12.5 + ([4.5 - 8.0] - 4.5)/0.2\mbox{  and  }\\
       [4.5] > 12.5
    \end{split}
\end{equation}
\begin{equation}
    \begin{split}
        AGN = AGN_1 \mbox{  and  }\\
        [4.5] > 15 - 0.5([4.5 - 8.0] - 1\mbox{  and  }\\
        ([4.5] > 14 + 0.5([4.5 - 8.0] - 2))\mbox{  and  }\\
       [4.5] > 15
    \end{split}
\end{equation}

We selected 1,117 sources that were likely PAH galaxy contaminants satisfying either of the following criteria:

\begin{equation}
    \begin{split}
        PAH_1 = [4.5 - 5.8] < (2.5/2)([5.8 - 8.0] - 1) \mbox{  and  }\\
        [4.5 - 5.8] < 1.55\mbox{  and  }\\
       [5.8 - 8.0] > 1\mbox{  and  }\\
       [4.5] > 11.5
    \end{split}
\end{equation}

\begin{equation}
    \begin{split}
        PAH_2 = [3.6 - 5.8] < (3.2/3)([4.5 - 8.0] - 1) \mbox{  and  }\\
        [3.6 - 5.8] < 2.25\mbox{  and  }\\
        [4.5 - 8.0] > 1\mbox{  and  }\\
       [4.5] > 11.5
    \end{split}
\end{equation}

We removed 1 source selected as a knot of possible shocked emission satisfying:

\begin{equation}
    \begin{split}
       KNOT = [3.6 - 4.5] > 1.05 \mbox{  and  }\\
        [3.6 - 4.5] > (1.2/0.55)([4.5 - 5.8] - 0.3) + 0.8\mbox{  and  }\\
        [4.5 - 5.8] \leq 0.85
    \end{split}
\end{equation}

We removed 1,234 sources with likely PAH aperture contamination according to the following criteria:

\begin{equation}
    \begin{split}
       PA=[3.6-4.5] - \sigma_{12} \leq 1.5\cdot([4.5-5.8]-\sigma_{23}-1) \mbox{  and  }\\
        [3.6 - 4.5] - \sigma_{12} \leq 0.4
    \end{split}
\end{equation}

After removing a combined 2,325 background sources, we were left with 486 candidates for YSO selection with photometry in bands 1 and 2 and either band 3 and/or 4. We calculated extinction coefficients for these sources and removed 5 sources with invalid values. 

We selected 45 YSOs using the following four-color IRAC cuts:
\begin{equation}
    \begin{split}
        [5.8-8.0]\geq 0.3+\sigma_{34} \mbox{  and  }\\
        [5.8-8.0]\leq 2 - \sigma_{34} \mbox{  and  }\\
        [3.6 - 4.5] \geq 0.2 + \sigma_{12} \mbox{  or  }\\
        [5.8-8.0] \leq 2.5-\sigma_{34} \mbox{  and  }\\
        [3.6 - 4.5] \geq 0.5 + \sigma_{12} \mbox{  and  }\\
        [5.8-8.0]>2 +\sigma_{34}
    \end{split}
\end{equation}

And 57 using the following additional four-color IRAC cuts:
\begin{equation}
    \begin{split}
        [3.6-5.8]\geq 0.5 + \sigma_{13} \mbox{  and  }\\
        [4.5-8.0]\geq 0.5 + \sigma_{24} \mbox{  and  }\\
        [3.6-4.5]\geq 0.15 + \sigma_{12} \mbox{  and  }\\
        [3.6-5.8]+\sigma_{13} \leq (0.14/0.04)\times ([4.5-8.0] - \sigma_{24}-0.5) +0.5 
    \end{split}
\end{equation}

For the 1,361 sources with detections at 5.8~\micron\ but not 8.0~\micron\ we applied the following cuts to select 8 YSOs:
\begin{equation}
    \begin{split}
        [3.6-4.5]-\sigma_{12}>0.3 \mbox{  and  }\\
        [4.5-5.8]-\sigma_{23}>0.3
    \end{split}
\end{equation}
 
We also imposed a final selection for sources with a large color excess, selecting 20 sources:
\begin{equation}
[Ks - 8.0] - \sigma_{k4} > 3
\end{equation}

Altogether, we identified 60 total YSOs through these selections, 9 of which were not identified using the 2MASS+IRAC, \textit{WISE},  \textit{WISE}+IRAC, or MMIRS+IRAC cuts described in Appendices~\ref{appendix:2MASSIRAC}, \ref{appendix:WISE}, \ref{appendix:WISEIRAC}, and \ref{appendix:MMIRSIRAC}.

\section{MIPS YSO Selection}\label{appendix:MIPS}
\setcounter{equation}{0}
We selected 39 YSOs using the MIPS 24~\micron\ photometry. We removed background sources following the criteria in Appendix~\ref{appendix:IRAC34} and then followed the methods of \citet{Winston2019, Winston2007} and  \citet{Gutermuth2008b} and imposed these selection criteria:
\begin{equation}
    \begin{split}
        [8.0 - 24] \geq1.0 + \sigma_{4m} \mbox{  and  }
        [5.8 - 8.0]\geq -0.1- \sigma_{34} \mbox{  or  }\\
        [8.0 - 24] \geq 0.6 + \sigma_{4m} \mbox{  and  } 
        [5.8 - 8.0]\geq 0.2 - \sigma_{34}
    \end{split}
\end{equation}

\begin{equation}
    \begin{split}
        [8.0 - 24] \geq 1.0 + \sigma_{4m} \mbox{  and  }
        [3.6-4.5]  \geq -0.1 - \sigma_{12} \mbox{  or  }\\
        [8.0 - 24] \geq 0.6 + \sigma_{4m} \mbox{  and  }
        [3.6 - 4.5]\geq 0.2 - \sigma_{12}
    \end{split}
\end{equation}

Of the remaining 39 YSOs, 10 were not detected using the 2MASS+IRAC, \textit{WISE},  \textit{WISE}+IRAC, or MMIRS+IRAC cuts described in Appendices~\ref{appendix:2MASSIRAC}, \ref{appendix:WISE}, \ref{appendix:WISEIRAC}, and \ref{appendix:MMIRSIRAC}.
\begin{deluxetable*}{|c|c|c|c|c|c|c|c|c|c|}
%\rotate
\tabletypesize{\scriptsize}
\tablecaption{SExtractor parameters for the IRAC long (L) and short (S) and MIPS mosaics\label{tab:sextractor}}
\tablewidth{0pt}
\tablehead{
\colhead{Parameters} & \colhead{3.6 \micron\ L} & \colhead{3.6 \micron\ S} & \colhead{4.5 \micron\ L} &\colhead{4.5 \micron\ S} & \colhead{5.8 \micron\ L} & \colhead{5.8 \micron\ S} & \colhead{8.0  \micron\ L} &\colhead{8.0 \micron\ S} &  \colhead{MIPS 24 \micron}
}
\startdata
{DETECT\_MINAREA} (pix)       & 3        & 3        & 3         & 3     & 3     & 3     & 3     & 3     & 3       \\
{DETECT\_THRESH }($\sigma$)   & 2.25     & 2.25     & 2.25      & 2.25  & 2.75  & 2.75  & 2.50  & 2.50  & 2.25    \\
{ANALYSIS\_THRESH} ($\sigma$) & 2.25     & 2.25     & 2.25      & 2.25  & 2.75  & 2.75  & 2.50  & 2.50  & 2.25     \\
{DEBLEND\_NTHRESH}            & 32       & 32       & 32        & 32    & 32     & 32     & 32     & 32     & 32       \\     
{DEBLEND\_MINCOUNT }          & 0.005    & 0.005    & 0.005     & 0.005 & 0.005     & 0.005     & 0.005     & 0.005     & 0.005       \\    
{PHOT\_APERTURES} (pix)       & 5        & 5        & 5         & 5     & 5     & 5     & 5     & 5     & 5       \\     
{PHOT\_AUTOPARAMS} (pix)      & 2.5, 3.5 & 2.5, 3.5 & 2.5, 3.5  & 2.5, 3.5  & 2.5, 3.5     & 2.5, 3.5     & 2.5, 3.5     & 2.5, 3.5     & 2.5, 3.5   \\ 
{MAG\_ZEROPOINT}              & 17.68169 & 17.68169 & 17.17799  & 17.17799  & 16.49360     & 16.49360     & 15.47308     & 15.47308     & 10.94358   \\
{SEEING\_FWHM} ('')           & 1.6      & 1.6      & 1.6       & 1.6   & 1.6     & 1.6     & 1.6     & 1.6     & 1.6       \\
{BACK\_SIZE }(pix)            & 4        & 10        & 4        & 10    & 3     & 3     & 2     & 2     & 2       \\
{BACK\_FILTERSIZE} (pix)      & 3        & 7        & 3         & 7     & 3     & 3     & 3     & 3     & 3       \\
\enddata
\end{deluxetable*}
\begin{deluxetable*}{|c|c|c|c|c|c|}
%\rotate
\tabletypesize{\scriptsize}
\tablecaption{SExtractor parameters for the MMIRS mosaics from SAO-8-19c and SAO-12-21a\label{tab:sextractormmirs}}
\tablewidth{0pt}
\tablehead{
\colhead{Parameters} & \colhead{J SAO-8-19c} & \colhead{H SAO-8-19c} & \colhead{K SAO-8-19c} &\colhead{J SAO-12-21a} & \colhead{H SAO-12-21a}
}
\startdata
{DETECT\_MINAREA} (pix)       & 3        & 3        & 3         & 3     & 3 \\
{DETECT\_THRESH }($\sigma$)   & 2.00     & 2.00     & 2.00      & 2.00  & 2.00 \\
{ANALYSIS\_THRESH} ($\sigma$) & 2.00     & 2.00     & 2.00      & 2.00  & 2.00  \\
{DEBLEND\_NTHRESH}            & 32       & 32       & 32        & 32    & 32 \\     
{DEBLEND\_MINCOUNT }          & 0.005    & 0.005    & 0.005     & 0.005 & 0.005 \\    
{PHOT\_APERTURES} (pix)       & 5        & 5        & 5         & 5     & 5  \\     
{PHOT\_AUTOPARAMS} (pix)      & 2.5, 3.5 & 2.5, 3.5 & 2.5, 3.5  & 2.5, 3.5  & 2.5, 3.5  \\ 
{MAG\_ZEROPOINT}              & 17.074   & 17.074   & 17.074    & 17.074  & 17.074  \\
{SEEING\_FWHM} ('')           & 1.6      & 1.6      & 1.6       & 1.6   & 1.6 \\
{BACK\_SIZE }(pix)            & 4        & 4        & 4        & 4    & 4   \\
{BACK\_FILTERSIZE} (pix)      & 3        & 3        & 3         & 3     & 3   \\
{Average calibration slope} & 0.033        & 0.089        & 0.035         & 0.035     & 0.066   \\
{Average zero point\tablenotemark{a}} & 23.664        & 24.431        & 23.184         & 24.200     & 24.398
\enddata
\tablenotetext{a}{While we ran the source extraction with an arbitrary value for MAG\_ZEROPOINT, we calculated the average zero point afterwards by taking the grand mean of the difference in 2MASS and MMIRS magnitudes over all the images in a sample. This value reflects the approximate zero point magnitude for MMIRS without accounting for the nonlinearity.}
\end{deluxetable*}



\end{document}